\documentclass[prb,aps,twocolumn,showpacs,amsmath,amsfonts,superscriptaddress]{revtex4-2}

\usepackage{graphicx,color}
\usepackage[english]{babel}
\usepackage{amsmath,amssymb}
\usepackage{stmaryrd,dsfont,txfonts}
\usepackage[sort&compress]{natbib}
\usepackage{subfigure}
\usepackage{mathrsfs}
\usepackage{colordvi}
\usepackage{epsfig}
\usepackage{float}
\usepackage{dcolumn}
\usepackage{ifpdf}

\usepackage{physics}

\usepackage{hyperref}
\hypersetup{
	colorlinks,%
	citecolor=blue,%
	filecolor=blue,%
	linkcolor=blue,%
	urlcolor=blue
}

\usepackage{bm} 

\newcommand{\kp}{\ensuremath{\bm{k}\cdot\bm{p}} }

\begin{document}

\title{Spin-orbit coupling in wurtzite quantum wells}

\author{J. Y.\ Fu}
\thanks{These authors contributed equally to this work.}
\affiliation{Department of Physics, Qufu Normal University, Qufu, Shandong, 273165, China}
\affiliation{Instituto de F\'{\i}sica de S\~ao Carlos, Universidade de S\~ao Paulo, 13560-970 S\~ao Carlos, S\~ao Paulo, Brazil}
\author{P. H. Penteado}
\thanks{These authors contributed equally to this work.}
\affiliation{Instituto de F\'{\i}sica de S\~ao Carlos, Universidade de S\~ao Paulo, 13560-970 S\~ao Carlos, S\~ao Paulo, Brazil}
\author{D. R. Candido}
\affiliation{Instituto de F\'{\i}sica de S\~ao Carlos, Universidade de S\~ao Paulo, 13560-970 S\~ao Carlos, S\~ao Paulo, Brazil}
\affiliation{Department of Physics and Astronomy, University of Iowa, Iowa City, Iowa 52242, USA}
\affiliation{Pritzker School of Molecular Engineering, University of Chicago, Chicago, Illinois 60637, USA}
\author{G. J. Ferreira }
\affiliation{Instituto de F\'{\i}sica, Universidade Federal de Uberl\^andia, Uberl\^andia, Minas Gerais 38400-902, Brazil}
\author{D. P. Pires}
\affiliation{Instituto de F\'{\i}sica de S\~ao Carlos, Universidade de S\~ao Paulo, 13560-970 S\~ao Carlos, S\~ao Paulo, Brazil}
\affiliation{Departamento de F\'isica Te\'orica e Experimental, Universidade Federal do Rio Grande do Norte, 59072-970 Natal, Rio Grande do Norte, Brazil}
\author{E. Bernardes}
\affiliation{Instituto de F\'{\i}sica de S\~ao Carlos, Universidade de S\~ao Paulo, 13560-970 S\~ao Carlos, S\~ao Paulo, Brazil}
\author{J. C. Egues}
\affiliation{Instituto de F\'{\i}sica de S\~ao Carlos, Universidade de S\~ao Paulo, 13560-970 S\~ao Carlos, S\~ao Paulo, Brazil}

\date{\today}

\begin{abstract}
Effective spin-orbit (SO) Hamiltonians for conduction electrons in wurtzite heterostructures are lacking in the literature, in contrast to zincblende structures. Here we address this issue by deriving such an effective Hamiltonian valid for quantum wells, wires, and dots with arbitrary confining potentials and external magnetic fields. We start from an 8$\times$8 Kane model accounting for the $s$--$p_z$ orbital mixing important to wurtzite structures, but absent in zincblende, and apply both quasi-degenerate perturbation theory (L\"owdin partitioning) and the folding down approach to derive an effective 2$\times$2 electron Hamiltonian. Focusing on wurtzite quantum wells, we later on also perform a self-consistent Poisson-Schr\"odinger calculation in the Hartree approximation to determine the relevant SO couplings. We obtain the usual $k$-linear Rashba term arising from the structural inversion asymmetry of the wells and, differently from zincblende structures, a bulk Rashba-type term induced by the inversion asymmetry of the wurtzite lattice. Our results show this latter term to be the main contribution to the Rashba coupling in wurtzite wells. We also find  linear- and cubic-in-momentum Dresselhaus contributions. Both the bulk Rashba-type term and the Dresselhaus terms originate exclusively from the admixture of $s$- and $p_z$-like states in wurtzites structures. Interestingly, in these systems the linear Rashba and the Dresselhaus terms have the same symmetry and can in principle cancel each other out completely, thus making the spin a conserved quantity. We determine the intrasubband (intersubband) Rashba $\alpha_\nu$ ($\eta$) and linear Dresselhaus $\beta_\nu$ ($\Gamma$) SO strengths of GaN/AlGaN single and double wells with one and two occupied subbands ($\nu=1,2$). For the GaN/Al$_{0.3}$Ga$_{0.7}$N single well with one occupied  subband, we obtain the total spin splitting coefficient $\alpha^{\rm eff}_{1}=\alpha_1+\beta_1\sim7.16$~meV$\cdot$\AA, in agreement with weak antilocalization measurements. In the case of two occupied subbands, we observe that the intersubband Rashba $\eta$ is much weaker than the intrasubband coupling $\alpha_\nu$. For double wells even in the presence of strong built-in electric fields (spontaneous and piezoelectric, crucial in GaN/AlGaN wells), we find a \emph{seemingly} symmetric potential configuration at which both the Rashba $\eta$ and Dresselhaus $\Gamma$ intersubband couplings exhibit their highest strengths. On the other hand, we observe that the intrasubband Dresselhaus coefficients $\beta_1$ and $\beta_2$ interchange their values as the gate voltage $V_g$ varies across zero; a similar behavior, though less pronounced, is seen for the Rashba couplings $\alpha_1$ and $\alpha_2$. We believe our general effective Hamiltonian for electrons in wurtzite heterostructures put forward here, should stimulate additional theoretical works on wurtzite quantum wells, wires, and dots with variously defined geometries and external magnetic fields.
\end{abstract}

\maketitle

\section{Introduction}

The spin-orbit interaction couples the electron spin and its momentum. While in atomic systems this relativistic effect arises from the Coulomb interaction within the atom, in mesoscopic semiconductor heterostrutures such as quantum wells, wires, and dots, the SO interaction originates from the  interplay of the confining, doping, Hartree, and external gate potentials. Unlike atomic systems, the SO coupling strength in these systems can be electrically controlled, thus providing a unique handle for the manipulation of the magnetic moment of the electron. Spin manipulation via the SO interaction is an important resource in spintronic devices and quantum information processing with spin qubits~\cite{Awschalom:2002,zutic:2004}.

\begin{figure}[htb!]
	\includegraphics[width=8.cm]{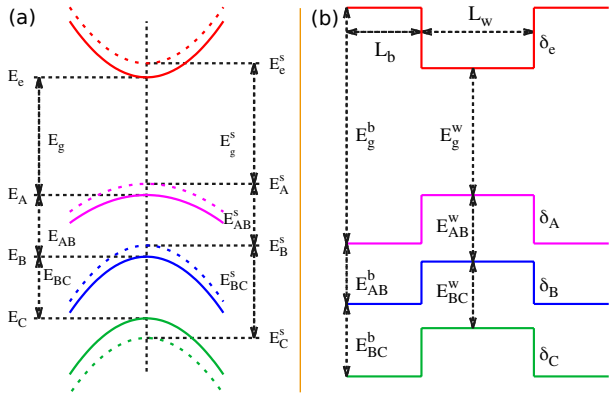}
	\caption {(a) Schematic of the dispersion relation (Chuang and Chang  basis set) of an unstrained (solid curves)
		and a strained (dashed curves) wurtzite semiconductor close to the $\Gamma$ point ($\bm{k}=0$). The superscript $\rm{s}$ in the band parameters indicates the strained case.  (b) Potential profile of a wurtzite single well of width $L_w$ and barriers of width $L_b$. The superscripts $b$ and $w$ stand for barrier and well, respectively. 
		No $s$--$p_z$ mixing is considered here since its effect is negligible on the band edges at the $\Gamma$ point.
	}
	\label{fig1}
\end{figure}

Spin orbit effects also underlie novel topological transport phenomena in diverse fields of quantum condensed matter such as topological insulators~\cite{KaneTI,bernevig2:2006}, Majorana fermions~\cite{Alicea,lutchyn:2010,oreg:2010}, and Weyl semimetals~\cite{weng:2015}. Recent proposals for stretchable~\cite{dettwiler:2014} spin helix~\cite{schliemann:2003, bernevig:2006,koralek:2009,salis:2012} and persistent skyrmion lattice excitations~\cite{PSL} in ordinary GaAs wells also highlight the important role of the SO in quantum wells.
 
So far, detailed theoretical and experimental studies on the SO coupling in semiconductors have been performed mostly in crystals with zincblende structure, including both bulk and confined systems~\cite{Winkler,Fu2015}. It is well established by now that the structural inversion asymmetry (SIA) of wells lead to a liner-in-momentum Rashba term~\cite{Bychkov} while the bulk inversion asymmetry (BIA) of the underlying zincblende crystal lattice gives rise to linear- and cubic-in-momentum Dresselhaus terms~\cite{Dresselhaus}. Wells with two subbands also have \textit{intersubband} Rashba- and Dresselhaus-like terms~\cite{Fu2015,Esmerindo1,Esmerindo2,PSL}. The interplay of the linear Rashba and Dresselhaus terms gives rise to some of the interesting phenomena mentioned in the previous paragraph in zincblende matrices.

On the other hand, in wurtzite structures, the presence of a hexagonal $c$ axis along the $z||(0001)$ direction allows for a linear BIA Rashba term~\cite{Rashba,Bihlmayer} in addition to a cubic BIA Dresselhaus term~\cite{Wang,Fu}. Besides these BIA terms, a usual SIA Rashba-like contribution~\cite{Bychkov} is also present in wurtzite wells as we shall see.

The SO coupling in wurztite semiconductors has attracted interest both experimentally~\cite{Weber,Schmult,Thillosen1,Kurdak,Lo1,Belyaev,Olshanetsky,cheng08,Lisesivdin,Stefanowicz} and 
theoretically~\cite{Voon,Wang,Fu,Litvinov1}. Experimentally, spin splitting energies from $0$ up to $13$ meV were reported in GaN-based heterostructures~\cite{Weber,Lisesivdin,Belyaev,Olshanetsky,Lo1}. Weak antilocalization measurements provide a SO splitting parameter value ranging from 5.5 to 10.01 meV$\cdot${\AA}~\cite{Schmult,Thillosen1,Kurdak,Belyaev,cheng08,Stefanowicz}.
Compared with these measurements, the value extracted from the beating pattern of Shubnikov-de Haas oscillations is very large, around $65$ meV$\cdot${\AA}, which is attributed to the structural inhomogeneity of the crystals~\cite{Thillosen2,Thillosen3}.

Theoretically, Lew Yan Voon \emph{et al.} investigated the linear Rashba-type term in bulk wurtzite semiconductors~\cite{Voon}
and found that the linear term is mainly determined by the mixing of the $s$ orbital of the conduction band and the $p_z$ orbital of the valence band ($s$--$p_z$ mixing)~\cite{RashbaI}. Wang \emph{et al.} studied the bulk cubic Dresselhaus SO interaction and demonstrated the existence of a zero Dresselhaus spin splitting surface in wurtzite semiconductors~\cite{Wang}.
Fu and Wu evaluated the bulk Dresselhaus coefficient in GaN, 0.32 eV$\cdot$\AA$^3$~\cite{Fu}, which has been experimentally verified via the circular photogalvanic measurement~\cite{Yin}.
More recently, Faria Junior \emph{et al.}~\cite{Paulo} investigated the bulk spin-orbit coupling effect in the wurtzite phase via \emph{ab initio} band structure calculations. 

In addition to studies on the bulk wurtzite structure, the SO coupling parameter in GaN/AlN  heterostructures, around 8 meV$\cdot${\AA}, was determined by Majewski via first-principles calculations~\cite{Majewski}.
Litvinov investigated the spin splitting of GaN/AlGaAs heterostructure \cite{Litvinov1} and GaN/InGaN quantum wells~\cite{Litvinov2}, with one occupied electronic subband. Following the basic framework of his formulation, Li \emph{et al.} determined the Rashba couplings associated with two occupied electronic subbands in GaN/Al$_{0.3}$Ga$_{0.7}$N wells, 
$\alpha_1 \approx \alpha_2 \sim 0.5$ meV$\cdot$\AA.

Although several investigations have been conducted, a comprehensive theory on the SO coupling in wurtzite crystals, from bulk to confined structures, is still lacking in the literature.
Moreover, for the available reports on the SO coupling, the $s$--$p_z$ orbital mixing, which we find is crucial in obtaining certain SO terms, was not taken into account in the Kane models used.  In addition, in the derivation of the effective electron Hamiltonian, the renormalization of the conduction band spinor component was not considered~\cite{Litvinov1}, thus leading to an energy dependent Schr\"odinger-type equation.

In this work, we account for these missing ingredients and establish a detailed systematic formulation for the electron SO interaction in wurtzite heterostructures. We then consider GaN/AlGaN wells, both single and double, involving the electron occupancy of either one or two subbands, and self-consistently determine the intrasubband (intersubband) Rashba $\alpha_\nu$ ($\eta$), $\nu=1,2$, and Dresselhaus $\beta_\nu$ ($\Gamma$) terms. By using an external gate voltage $V_{\rm g}$, we also discuss the electrical control of all relevant SO couplings. These SO terms are helpful to investigate spin related properties in semiconductors with wurtzite phase, especially in confined wurtzite nanostructures (wells, wires, and dots).

More specifically, we use the basis set defined by Chuang and Chang \cite{Chuang} (``CC basis'') to construct an 8$\times$8 Kane model, in which we account for the $s$--$p_z$ orbital mixing (see Appendix~\ref{app:grouptheory}). The CC basis (Table~\ref{table:basiscc}) is a solution of the bulk Hamiltonian at $\bm{k}=0$ ($\Gamma$ point) with the k-independent SO coupling being partially included [Eq.~\eqref{eq:h0cc}]. Therefore, the Kane model contains nonzero off-diagonal elements even at $\bm{k}=0$ [see Eq.~\eqref{eq:kanebcc}], implying  that the diagonal elements of the Kane matrix in general do not describe the actual band edge energies, Fig.~\ref{fig1}.
We also derive the bulk Kane Hamiltonian in the ``diagonal basis'', in which the Kane model is diagonal at $\bm{k}=0$ (Appendix~\ref{app:diag}). This helps us determine the diagonal elements (not actual band edge energies) of the Kane model in the CC basis as well as the corresponding ``virtual" band offsets (see Table \ref{table:bandparameter}).  
Having the bulk Kane model at hand, we then construct its analogue for heterostructures.

We use quasi-degenerate perturbation theory (L\"owdin partitioning)~\cite{Lowdin}  and the folding down approach~\cite{Esmerindo1,Esmerindo2} to derive an effective 2$\times$2 electron Hamiltonian [Eq.\eqref{eq:well3}] from the 8$\times$8 Kane model for wurtzite heterostructures. As opposed to what has been reported in the literature~\cite{Litvinov1,li:2011,li2:2011}, we arrive at a \emph{genuine} Schr\"odinger-type equation, i.e., an energy-independent effective Hamiltonian, since we account for the renormalization of the conduction band spinor component (Appendix~\ref{app:folding}).

In addition to the usual linear in momentum Rashba term induced by the structural inverstion asymmetry of the wells, see Eqs.~\eqref{eq:etahcc} and \eqref{eq:etawcc}, we obtain a bulk Rashba-type SO coupling as a function of the $s$--$p_z$ orbital mixing [Eq.~\eqref{eq:etaccc}]. Within the eight bands considered ($\bm{s}$-conduction and $\bm{p}$-valence bands), the \kp interaction accounted for in the valence band subspace gives rise to the Dresselhaus coupling (Appendix~\ref{app:hvq}). This is in contrast to systems with the zincblende structure, in which the Dresselhaus term is only present if the \kp interaction with remote bands, e.g., the $\bm{p}$-conduction band, is included~\cite{Cardona}. We also derive a general  effective electron Hamiltonian containing all relevant SO terms for wurtzite nanostructures in the presence of a magnetic field and an arbitrary 3D confinement [see Eqs.~\eqref{eq:heff3}--\eqref{eq:heff3so}].

Based on our results for the SO terms, we self-consistently calculate the total spin splitting coefficient for a GaN/AlGaN single well with one occupied subband $\alpha_{1}^{\rm eff}=\alpha_1+\beta_1\sim 7.16$ meV$\cdot$\AA~[Figs.~\ref{fig3}--\ref{fig5}], in agreement with weak antilocalization measurements~\cite{Stefanowicz,Belyaev,Kurdak}. We determine as well the several distinct terms composing the Rashba SO coupling: the ``bulk''~\cite{footnote-bulk}, the Hartree, and the structural well contributions [Eq.~\eqref{eq:etadia}], and show that the strength of the SO coupling follows from the interplay of all these components, Fig.~\ref{fig4}. For a similar calculation for zincblende quantum wells see Refs.~\cite{Esmerindo1,Esmerindo2}.
We note that the bulk Rashba term dominates over all the other contributions. Accordingly, the latter remains essentially constant as a function of the gate voltage $V_{\rm g}$, Fig.~\ref{fig4}(a).
When the wells have two occupied subbands (higher electron density), we find that the intersubband Rashba strength $\eta$ is much weaker than the intrasuband coupling $\alpha_\nu$, Fig~\ref{fig6}.

On the other hand, for GaN/AlGaN double wells [Fig.~\ref{fig7}], interestingly, we find a \emph{seemingly} symmetric configuration depending on the relative ratio of the Al content between the central and lateral barriers, even in the presence of the strong built-in electric field (spontaneous and piezoelectric). At this configuration, a maximal strength of the intersubband Rashba $\eta$ and Dresselhaus $\Gamma$ couplings occurs. In addition, by varying $V_{\rm g}$ we observe that the Dresselhaus couplings $\beta_1$ and $\beta_2$ change dramatically and almost interchange their values. Although less pronounced, a similar behavior also holds for the Rashba couplings $\alpha_1$ and $\alpha_2$, see Fig.~\ref{fig8}.

The paper is organized as follows. In Secs.~\ref{sec:kanem}-\ref{sec:h2d}, we present the model and method used. Specifically, in Sec.~\ref{sec:kanem}, we review the \kp method and apply it to obtain an 8$\times$8 Kane model for both unstrained and strained wurtzite wells.
In Sec.~\ref{sec:eeh}, we  derive a 2$\times$2 3D SO Hamiltonian for electrons from the Kane model obtained in Sec.~\ref{sec:kanem}. In this derivation, we use both the L\"owdin partitioning method and the folding down approach. By projecting the 3D Hamiltonian onto the quantum well subbands (obtained in a self-consistent way),  we derive in Sec.~\ref{sec:h2d} an effective 2D electron Hamiltonian containing all the relevant SO terms. For concreteness, we present and discuss numerical results for GaN/AlGaN wells in Sec.~\ref{sec:algan}.
We summarize our main findings in Sec.~\ref{sec:summary}.

\section{8$\times$8 Kane model: bulk $\rightarrow$ heterostructures}
\label{sec:kanem}

\subsection{The \texorpdfstring{\kp}{kd} method: general remarks and notation}
\label{sec:kaneb}

For an electron on a microscopic periodic potential $V(\bm{r})$, the Schr\"odinger equation for the periodic part $u_{\nu\,\bm{k}}(\bm{r})$ of the Bloch function is given by \cite{Kane,Winkler,voon-book}
\begin{multline}
  \label{eq:schrodinger}
  \Bigg[ 
  	\frac{p^2}{2m_0}+V(\bm{r})+H_{\rm so}
  	+
  	\cfrac{\hbar}{m_0}\bm{k\cdot \bm{\pi}}
  \Bigg]u_{\nu\,\bm{k}}(\bm{r})
  \\
  = \Bigg(E_\nu(\bm{k})-\frac{\hbar^2k^2}{2m_0}\Bigg)u_{\nu\,\bm{k}}(\bm{r}),
\end{multline}
where $m_0$ is the bare electron mass, $\nu$ is a band index for each wave vector $\bm{k}$, $\bm{p}$ is the momentum operator, and 
\begin{equation}
 \bm{\pi} = \bm{p} + \frac{\hbar}{4m_0c^2}\bm{\sigma}\times\bm{\nabla}V(\bm{r}), 
 \end{equation}
 with $\bm{\sigma} = (\sigma_x, \sigma_y, \sigma_z)$ being the Pauli matrices. The spin-orbit coupling appears in Eq.~\eqref{eq:schrodinger} as the $\bm{k}=0$ term
\begin{align}
	\label{eq:hso}
	H_{\rm so} &= \frac{\hbar}{4m_0^2c^2}\bm{\nabla} V(\bm{r})\times\bm{p}\cdot\bm{\sigma},
\end{align}
and the $\bm{k}$-linear term $\dfrac{\hbar}{m_0}\bm{k\cdot \bm{\pi}}$.

The Hamiltonian above considers an unstrained crystal. For a finite strain, one must add the strain related couplings $H_{\rm strain}$ to Eq.~\eqref{eq:schrodinger} \cite{Bir,Bahder:1990}. We will discuss the strain tensor and its effects in Section \ref{sec:ccs}.

To solve Eq.~\eqref{eq:schrodinger} in the vicinity of the $\Gamma$ point ($\bm{k} \approx 0$) within the \mbox{\kp} approach, one must define a basis set $\{u_{\nu\,0}(\bm{r})\}$ at $\bm{k} = 0$ to expand $u_{\nu \bm{k}}(\bm{r})$. This choice of basis set is in principle not unique and leads to different representations of the Kane model~\cite{Winkler,voon-book}. Ideally, one would prefer to work on a basis set that diagonalizes Eq.~\eqref{eq:schrodinger} at $\bm{k} = 0$, as it is commonly done for zincblende structures. However, it is often more interesting to use a basis defined by the irreducible representations (IRREPs) of the crystal's group, which is not always diagonal at $\bm{k} = 0$. This is the case for the basis set defined by Chuang and Chang~\cite{Chuang} and introduced in the next section.

\subsection{8$\times$8 Kane model: bulk}
\label{subsec:kaneb}

In the following sections, we build an 8$\times$8 bulk Kane model (unstrained/strained) for wurtzite crystals using a slightly modified CC basis (primed basis CC$'$, see Appendix~\ref{app:grouptheory}), which incorporates previously neglected effects of the $s$--$p_z$ coupling~\cite{Voon}. 

\subsubsection{CC basis: unstrained case}
\label{sec:kaneccns}

Our basis set is defined by splitting the total Hamiltonian in Eq.~\eqref{eq:schrodinger} as $H = H_0^{\rm CC} + W^{\rm CC}(\bm{k})$, with

\begin{align}
\label{eq:h0cc}
H^{\rm CC}_0 &= \frac{p^2}{2m_0}+V(\bm{r})+H_{{\rm so}z},
\\
\label{eq:wkcc}
W^{\rm CC}(\bm{k}) &= \cfrac{\hbar}{m_0}\bm{k\cdot \bm{\pi}}+H_{{\rm so}x}+H_{{\rm so}y},
\\
H_{{\rm so}j} &= \cfrac{\hbar}{4m_0^2c^2}\Bigg(\bm{\nabla}
V(\bm{r})\times\bm{p}\Bigg)_j\,\sigma_j,
\end{align}
where $j = \{x, y, z\}$ labels the spin components of $H_{\rm so}$. The chosen CC basis is composed by the eigenstates that diagonalize $H_0^{\rm CC}$, which includes only the $z$-component ($H_{{\rm so}z}$) of the $\bm{k}=0$ spin-orbit interaction. Therefore, our Hamiltonian will not be diagonal at $\bm{k} = 0$, since the perturbation $W^{\rm CC}(\bm{k})$ contains the $\bm{k}$-independent $H_{{\rm so}x}$ and $H_{{\rm so}y}$ terms.

Wurtzite crystals comprise two interpenetrating hexagonal lattices that transform according to the space group $P6_3mc$ ($C_{6v}^4$), yielding the $C_{6v}$ point group at $\Gamma$ ($\bm{k}=0$). Hence, the solutions of $H_0^{\rm CC}$ are given by single-group states belonging to the $\Gamma_1$ ($S$ and $Z$) and $\Gamma_5$ ($\{X,Y\}$) IRREPs of $C_{6v}$, with well defined spin ($\uparrow$ and $\downarrow$) along the $z$ direction. Note that differently from zincblende crystals, the $z$ direction in the wurtzite unit cell is nonequivalent to the $x$ and $y$ directions. This allows for an $s$--$p_z$ mixing~\cite{RashbaI,Voon}, thus leading to the hybridized $S'$ and $Z'$ orbitals shown in Table \ref{table:basiscc}.

Using the CC basis listed in Table \ref{table:basiscc} [$|\nu' \rangle$ states in Table \ref{table:basiscc2}, Appendix~\ref{app:grouptheory}], we can now build the bulk 8$\times$8 Kane Hamiltonian $H_{8\times 8}^{\rm CC}$. We find~\cite{footnote-PT} (see Appendix~\ref{subsec:kp})
\begin{widetext}
	\begin{small}
		\begin{eqnarray}
		\label{eq:kanebcc}
		H^{\rm CC}_{8\times 8} = \cfrac{p^2}{2m_0}
		+
		\renewcommand{\arraystretch}{2.0}
		\begin{pmatrix}
		0 & 0 & -\frac{1}{\sqrt{2}}P_2k_+ & 0 &\frac{1}{\sqrt{2}}P_2 k_- & -i\sqrt{2}\Delta_{\rm sz} & P_1k_z & 0\\
		
		0 & 0 & 0 & \frac{1}{\sqrt{2}}P_2 k_-  & -i\sqrt{2}\Delta_{\rm sz}  & -\frac{1}{\sqrt{2}}P_2k_+ & 0 &P_1k_z \\
		
		-\frac{1}{\sqrt{2}}P_2 k_- &  0  &  -E_g^{\rm CC} & 0 & 0 & 0 & 0 & 0 \\
		
		0  & \frac{1}{\sqrt{2}}P_2k_+  & 0 &  -E_g^{\rm CC}   & 0 & 0 & 0 & 0 \\
		
		\frac{1}{\sqrt{2}}P_2k_+  & i\sqrt{2}\Delta_{\rm sz}  & 0 & 0 & -E_g^{\rm CC} - E_{AB} & 0 & 0 & \sqrt{2}\Delta_3 \\
		
		i\sqrt{2}\Delta_{\rm sz} & -\frac{1}{\sqrt{2}}P_2k_-  & 0 & 0 & 0 & -E_g^{\rm CC} - E_{AB} & \sqrt{2}\Delta_3 & 0  \\
		
		P_1k_z & 0 & 0 & 0 & 0 & \sqrt{2}\Delta_3  &  -E_g^{\rm CC} - E_{AC}  & 0     \\
		
		0 &P_1k_z  & 0 & 0 & \sqrt{2}\Delta_3  & 0 & 0 &  -E_g^{\rm CC} - E_{AC}
		\end{pmatrix}.
		\end{eqnarray}
	\end{small}
\end{widetext}

\begin{table}[H]
	\caption{CC basis functions $u^{\rm CC}_{\nu\,\bm{0}}(\bm{r}) \equiv u_{\nu\,\bm{0}}(\bm{r}) \equiv \braket{\bm{r}}{\nu}$, with $\nu$=$1,2,...8$. The states $\ket{S^\prime}$ and $\ket{Z^\prime}$ transform like scalars and $\{\ket{X},\ket{Y}\}$ transform like vectors. The corresponding doubly-degenerate band edges (at the $\Gamma$ point) are denoted by $E_e$ (conduction) and $E_A$, $E_B$, $E_C$ (valence) bands, respectively. The $s$--$p_z$ mixed orbitals are $\ket{S'}=q_s\ket{S}+q_z\ket{Z}$, and $\ket{Z'} = q_s\ket{Z}-q_z\ket{S}$, with $q_s^2+q_z^2 = 1$ and $q_z \ll 1$ (Appendix~\ref{app:grouptheory}).}
	\begin{ruledtabular}
		\begin{tabular}{cccc}
			& $\nu$ & $\ket{\nu}$ & $C_{6v}$ IRREP  \\ \hline
			$e$ & 1 & $\ket{iS^\prime\uparrow}$ & $\Gamma_1$  \\
			$e$ & 2 & $\ket{iS^\prime\downarrow}$ & $\Gamma_1$ \\
			$A$ & 3 & $-\frac{1}{\sqrt 2}\ket{X+iY\uparrow}$ & $\Gamma_5$ \\
			$A$ & 4 & $+\frac{1}{\sqrt 2}\ket{X-iY\downarrow}$ & $\Gamma_5$ \\
			$B$ & 5 & $+\frac{1}{\sqrt 2}\ket{X-iY\uparrow}$ & $\Gamma_5$ \\
			$B$ & 6 & $-\frac{1}{\sqrt 2}\ket{X+iY\downarrow}$ & $\Gamma_5$ \\
			$C$ & 7 & $\ket{Z^\prime\uparrow}$ & $\Gamma_1$ \\
			$C$ & 8 & $\ket{Z^\prime\downarrow}$ & $\Gamma_1$ \\
		\end{tabular}
	\end{ruledtabular}
	\label{table:basiscc}
\end{table}

The nonzero matrix element $\Delta_{\rm sz}$ was neglected in the original work by Chuang and Chang~\cite{Chuang}, see Appendix~\ref{app:grouptheory} for details on the origin of this term. However, it is allowed by symmetry as mentioned above. Namely, this spin-orbit coupling reads 
\begin{align}
	\label{eq:deltasz}
	\Delta_{\rm sz} =& \dfrac{\hbar^2}{4m_0^2c^2}\mel{Y}{\dfrac{\partial V}{\partial y} \dfrac{\partial}{\partial z} - \dfrac{\partial V}{\partial z} \dfrac{\partial}{\partial y}}{S'}.
\end{align}
Even though this term has a negligible effect on the band edges at $\bm{k}=0$, it is crucial in obtaining the relevant SO terms at $\bm{k}\neq 0$.

The diagonal matrix elements in~(\ref{eq:kanebcc}) are the eigenenergies of $H^{\rm CC}_0$ [see Eq.~\eqref{eq:h0cc}]; here we set $E_e \equiv 0$ (energy reference), $E_A=-E_g^{\rm CC}$, $E_B=-E_g^{\rm CC}-E_{AB}$, and $E_C=-E_g^{\rm CC}-E_{AC}$, with the subscripts $e$ denoting the lowest conduction band and $A$, $B$, and $C$ the topmost three valence bands, respectively. The term $E_g^{\rm CC} \approx E_g$ corresponds to the band gap (see discussion below). The energy differences between the valence band edges are given by
\begin{equation}
	\label{eq:eabccc}
	E_{AB} = 2\Delta_2, ~~~~ E_{AC} = \Delta_1+\Delta_2,
\end{equation}
where $\Delta_1\equiv\Delta_{\rm cr}$ is the crystal-field splitting energy, and $\Delta_{2}$ and $\Delta_{3}$ are two SO split-off energy parameters, which read
\begin{align}
	\Delta_{2} &= -\dfrac{\hbar^2}{4m_0^2c^2}\bra{Y}\dfrac{\partial V}{\partial x} \dfrac{\partial}{\partial y} - \dfrac{\partial V}{\partial y} \dfrac{\partial}{\partial x}\ket{X},
	\\
	\nonumber
	\Delta_{3} &= \frac{\hbar^2}{4m_0^2c^2}\bra{Y}\frac{\partial V}{\partial y} \frac{\partial}{\partial z} - \frac{\partial V}{\partial z} \frac{\partial}{\partial y}\ket{Z}.
\end{align}
These are commonly assumed to be related by $\Delta_2=\Delta_3\equiv\Delta_{\rm so}/3$~\cite{Ren}, following the quasi-cubic approximation~\cite{voon-book}. The off-diagonal $\bm{k}$-linear terms are defined by Kane parameters $P_1 = -(i\hbar/m_0)\bra{S^\prime}p_z\ket{Z}$ and
$P_2 = -(i\hbar/m_0)\bra{S^\prime}p_x\ket{X}=-(i\hbar/m_0) \bra{S}p_y\ket{Y}$, with $k_{\pm}=k_x \pm ik_y$.

The Hamiltonian in Eq.~\eqref{eq:kanebcc} is well defined in terms of the matrix elements shown above. Note, however, that the diagonal matrix elements do not correspond to the real band edges, since $H^{\rm CC}_{8\times 8}$ is not diagonal at $k=0$. Nevertheless, we can  safely use $E_g^{\rm CC} \approx E_g$ (real band gap) since (i) Eq.~\eqref{eq:kanebcc} shows that the topmost valence band $A$ does not couple to other bands at $\bm{k}=0$, and (ii) the $\Delta_{\rm sz}$ coupling between the conduction and valence bands leads to a negligible second-order correction for the conduction band at $\bm{k}=0$, i.e., {$E_g = E_g^{\rm CC} - 2\Delta_{\rm sz}^2/(E_g^{\rm CC}+E_{AB}) \approx E_g^{\rm CC}$}. 

In the following, we derive the bulk Rashba term (see Appendix~\ref{subsec:kp} for more details). 

\subsubsection{Bulk Rashba dispersion}
\label{sec:rashbadis}
The conduction band of wurtzite crystals has intrinsic Rashba-like spin textures~\cite{Rashba}; this interesting feature was associated with the $s$--$p_z$ mixing~\cite{Voon}. To obtain a closed expression for this spin-orbit coupling within our model, we use L\"owdin perturbation theory to derive an effective $2\times2$ bulk Hamiltonian for the conduction band. Up to fourth order, we obtain

\begin{equation}
	H^{\rm CC}_{2\times 2} =
	\dfrac{\hbar^2k_{z}^2}{2m_{\bot}}+\dfrac{\hbar^2k_{\|}^2}{2m_{\|}}+\alpha_{\rm bulk}\left(\sigma_x k_y-\sigma_y k_x\right),
\end{equation}
with $k^2_{\|}=k_x^2+k_y^2$, effective masses
\begin{equation}
\label{eq:effmassz}
   \dfrac{1}{m_{\bot}}=\dfrac{1}{m_0}+\dfrac{2P^2_1}{\hbar^2}\cfrac{1}{E_g+\Delta_1+\Delta_2},
\end{equation}
\begin{equation}
\label{eq:effmassp}
   \dfrac{1}{m_{\|}}=\dfrac{1}{m_0}+\dfrac{P^2_2}{\hbar^2}\left(\dfrac{1}{E_g}+\dfrac{1}{ E_g+2\Delta_2}\right),
\end{equation}
and Rashba coupling
\begin{multline}
\label{eq:rashbab}
\alpha_{\rm bulk} = \dfrac{2 P_2 \Delta_{\rm sz}}{(E_g + 2\Delta_2)} 
+ \dfrac{4 P_2 \Delta _3^2 \Delta _{\rm sz}}{(E_g+2\Delta_2)^2 (E_g+\Delta _1+\Delta _2)}
 \\
- \dfrac{8 P_2 \Delta_{\rm sz}^3}{(E_g+2\Delta_2)^3}.
\end{multline}

Note indeed that $\alpha_{\rm bulk} \propto \Delta_{\rm sz}$, which is finite only due to the broken cubic symmetry of the wurtzite crystal that allows for the $s$--$p_z$ mixing.

\subsubsection{CC basis: strained case}
\label{sec:ccs}
The application of an external stress on a bulk semiconductor leads to a shift of the energy levels and/or a splitting of the
heavy--light holes degeneracy~\cite{Bir}. We discuss now, how strain effects change the band edges of the \kp~Hamiltonian in Eq.~\eqref{eq:kanebcc}.

We restrict ourselves to the case of biaxial strain, i.e.,
\begin{eqnarray}
&&\varepsilon_{xx}=\varepsilon_{yy} \neq \varepsilon_{zz} \nonumber \\
&&\varepsilon_{xy}=\varepsilon_{yz}=\varepsilon_{zx}=0,
\end{eqnarray}
where $\varepsilon_{ij}$ ($i,j=x,y,z$) is the strain tensor. Notice that strain appearing in heterostructures is in general caused by a lattice mismatch at the interfaces. For a strained-layer wurtzite crystal pseudomorphically grown along the $z$ direction, the components of the strain tensor assume the following values: $\varepsilon_{xx}$=$\varepsilon_{yy}$=${(a_s-a)}/{a}$ and $\varepsilon_{zz}$=$-2{C_{13}}\varepsilon_{xx}/{C_{33}}$, where $a_s$ is the lattice constant of the substrate and $a$ of the epitaxy layer. The parameters $C_{13}$ and $C_{33}$ are the elastic stiffness constants~\cite{Chuang,Reed,Im,Vurgaftman}.

The conduction band edge has a hydrostatic energy shift given by $\Delta  E_e$=$a_{c_1}\varepsilon_{zz}$+ $a_{c_2}(\varepsilon_{xx}+\varepsilon_{yy}$), with $a_{c_1(c_2)}$ the conduction band deformation potential. The topmost $A$ valence band edge shifts according to $\Delta  E_{A}$=$S_1$+$S_2$, where $S_1$ and $S_2$ are written as $S_1$=$D_1\varepsilon_{zz}$+$D_2(\varepsilon_{xx}+\varepsilon_{yy})$ and $S_2$=$D_3\varepsilon_{zz}$+$D_4(\varepsilon_{xx}+\varepsilon_{yy})$, with $D_{1-4}$ the valence band deformation potentials~\cite{Chuang,Shan}.
Therefore, the band gap variation, i.e., $\Delta {E}_g=E_g^s-E_g$, is given by  $\Delta  E_e-\Delta  E_{A}$, with the superscript ``s'' denoting parameters in the strained case.
The energy differences between valence bands read
\begin{align}
  \label{eq:e97ws}
  E^s_{AB} &= E_{AB}=2\Delta_2, \\
  E^s_{AC} &= E_{AC}+S_2=\Delta_1+\Delta_2+S_2, 
  \label{eq:e97bws}
\end{align}
where one can explicitly see the band edge corrections due to the deformation potentials [cf. Eqs.~\eqref{eq:eabccc}, \eqref{eq:e97ws}, and \eqref{eq:e97bws}].

In Fig.~\ref{fig1}(a) we show the diagonal matrix elements for both cases: unstrained (solid curves) and strained (dashed curves). Here, the only effect caused by strain is a shift in the band edges; one can then straightforwardly write down the corresponding Kane Hamiltonian by simply replacing $E_g$, $E_{AB}$, and $E_{AC}$ in Eq.~\eqref{eq:kanebcc} by $E^s_g$, $E^s_{AB}$, and $E^s_{AC}$, respectively.

\subsubsection{Some remarks}

The Kane model in Eq.~\eqref{eq:kanebcc} was constructed including only the lowest conduction and top valence bands (see Table~\ref{table:basiscc}). This model only provides a good description of the electronic states in the conduction band. In order to properly describe holes, especially heavy and light holes, it is necessary to include additional bands~\cite{voon-book}.
Nevertheless, here we are just interested in the conduction band. From Eqs.~(\ref{eq:effmassz}) and~(\ref{eq:effmassp}), for instance, we can calculate the longitudinal and transversal electronic effective masses, respectively, which, as we shall see in Sec.~\ref{sec:algan}, are in excellent agreement with experimental results.

We also derive a Kane Hamiltonian using the ``diagonal basis'', in which the $8\times8$ matrix is diagonal at $\mathbf{k}=0$ (see Appenix~\ref{app:diag}). This helps us determine the diagonal elements (not actual band edge energies) of the Kane model in the CC basis, as well as the corresponding ``virtual" band offsets (see Table \ref{table:bandparameter}) to correctly describe heterostructures. Note, though, that the diagonal basis has a direct dependence on the band parameters.

\subsection{8$\times$8 Kane model: heterostructures}
\label{sec:kanewell}

We shall focus now on the Kane model for wurtzite heterostructures, more especifically quantum wells grown along the $z||$(0001) direction ($c$ axis), for which experimental data are available. Later on, we show a general formulation valid for wells, as well as wires and dots [see Sec.~\ref{subsec:general}].

Due to the different band edges at $\bm{k}=0$ for different materials, a sharp jump of the bands (offsets) happens at the interfaces, which introduces position-dependent potentials representing the different layers. For simplicity, we refer here to the unstrained case; the generalization to the strained case is straightforward (see Sec.~\ref{sec:ccs}).

\subsubsection{Unstrained quantum wells}

The Hamiltonian for wells looks exactly the same as that in Eq.~(\ref{eq:kanebcc}), except that now one has to replace $k_z \rightarrow -i(\partial/\partial z)$, and introduce in the diagonal $z$-dependent potentials to account for the band offsets. 
Note that the diagonal basis functions depend on the band parameters (see Table \ref{tabel:basisdia} in Appendix~\ref{app:diag}), and hence, have different values on the well and on the barriers. This may lead to unnecessary complications in practice~\cite{Bastard}. For convenience, we only consider the 8$\times$8 Kane model for wells written in the CC basis. We emphasize that, strictly speaking, even in the CC basis set, the basis functions are $z$-dependent, as the periodic part of the Bloch functions, i.e., ${S}$, ${X}$, ${Y}$ and ${Z}$, see Table \ref{table:basiscc}, can be different for the several layers. Here we neglect this difference, as it is usually done for zincblende heterostructures~\cite{Bastard,Burt}.
 
Let us first analyze the $z$-dependent structural potential added to the diagonal part of Hamiltonian~(\ref{eq:kanebcc}), for both the conduction and valence bands. From Fig.~\ref{fig1}(b), where we schematically show the band offsets, we can straightforwardly obtain
\begin{align}
  \label{eq:Ve}
   V_{w-e}(z) = \delta_e h_w(z), ~~ V_{w-i}(z) = - \delta_i h_w(z),\; i\in\{A,B,C\},
\end{align}
where $h_w(z)$=$\Theta(z-\frac{L_w}{2})+\Theta(-z-\frac{L_w}{2})$ describes a dimensionless square well profile, with $\Theta(z)$ the Heaviside function. In this expression, the center of the well has been taken as the origin in the $z$ direction ($z=0$). 

Once again, in addition to the $z$-dependent potentials in (\ref{eq:Ve}), the off-diagonal elements $\Delta_3$, $\Delta_{\rm sz}$, and the Kane parameters $P_1$ and $P_2$ should be $z$-dependent, since in principle they may have distinct values in different layers. In most wurtzite materials, though, they have very similar values~\cite{Vurgaftman}, and from now on, we use them as $z$-independent parameters.
 
The discussion above on the Kane Hamiltonian for a quantum well is based on the single-electron picture, in which the electrons experience only the structural potential of the well. Below we focus on modulation doped quantum wells in the Hartree approximation. In this case, besides the structural well potential, the doping (from dopants) potential, the pure electron Hartree (from electrons) potential, as well as the external gate potential also contribute to the total electron potential.

\subsubsection{Strained quantum wells}

For strained wurtzite heterostructures, an internal potential (built-in electric field), due to the strain-induced piezoelectric polarization and/or spontaneous polarization, usually plays a significant role as well~\cite{Ambacher}. To calculate the built-in fields, two types of boundary conditions are usually used: (i) periodic boundary conditions, i.e., the potentials of the external surfaces are equal, and (ii) neutral external surfaces~\cite{Vogl,Grandjean}.
The corresponding expressions for the electric fields $E_{w}$ (well) and $E_{b}$ (barriers) in both cases are given by

\begin{flushleft}
(i) periodic boundary conditions
\end{flushleft}
\begin{eqnarray}
\label{eq:eboun1}
E_{w}=\cfrac{2L_{b}(P_{b}-P_w)}{\epsilon_0\epsilon_r (L_w+2L_{b})}, ~~~ E_{b}=E_w+\cfrac{P_w-P_{b}}{\epsilon_0\epsilon_r},
\end{eqnarray}

\begin{flushleft}
(ii) neutral external surfaces
\end{flushleft}
\begin{eqnarray}
\label{eq:eboun2}
E_{w}=\cfrac{P_{b}-P_w}{\epsilon_0\epsilon_r},~~~E_{b}=0,
\end{eqnarray}
where $\epsilon_0$ is the vaccum permittivity and $P_{w}$ and $P_{b}$ are the polarization fields (spontaneous and piezoelectric) appearing in the well and barriers, respectively. Here we have assumed a uniform dielectric constant $\epsilon_r$ throughout the system. Note that, as the width of the barriers goes to infinity, the two types of boundary conditions become equivalent.

The direction and magnitude of the spontaneous polarization along the $c$ axis of a wurtzite crystal can be determined experimentally~\cite{Yu,Ambacher,Martin}.
On the other hand, the magnitude of the piezoelectric polarization reads
\begin{eqnarray}
\label{eq:pe}
P=2\cfrac{a_s-a}{a}\left(e_{31}-e_{33}\cfrac{C_{13}}{C_{33}}\right),
\end{eqnarray}
where $e_{31}$ and $e_{33}$ are piezoelectric coefficients. 
Both polarizations are calculated for the well and also for the barrier.

\subsubsection{Total potential}

The total self-consistent conduction and valence band potentials ${V}_{e}(z)$ and ${V}_i(z)$, $i\in\{A,B,C\}$, respectively, in strained wurtzite quantum wells are given by
\begin{equation}
\label{eq:vi}
  {V}_{e}(z)={V}_H(z)+{V}_{w-e}(z),~{V}_i(z)={V}_H(z)+{V}_{w-i}(z),
\end{equation}
with ${V}_H$ the Hartree potential, which also has several contributions, namely,
\begin{equation}
\label{eq:VHartree}
V_{H}(z) = {V}_{\rm elect}(z) + {V}_{\rm d}(z) + {V}_{\rm int}(z) + {V}_{\rm g}(z),
\end{equation}
where ${V}_{\rm elect}$ is the pure electron Hartree potential, ${V}_{\rm d}$ is the doping potential, ${V}_{\rm int}$ is the internal potential (due to the built-in electric field), and ${V}_{\rm g}$ is the external gate potential (see Appendix~\ref{app:potential} for details).
 
\section{Effective 3D Hamiltonian}
\label{sec:eeh}
Based on the 8$\times$8 Kane model for wurzite quantum wells discussed in the last section, we now derive an effective 2$\times$2 Hamiltonain for the conduction band.
To this end we can either use the L\"owdin partitioning method, discussed in detail in Refs.~\cite{Lowdin,Winkler}, or the folding down approach, shown in Appendix~\ref{app:folding}. We use both approaches and obtain the same result. We also derive the effective model for wells using the theory of  invariants, as shown in Appendix~\ref{app:invariant}.

\subsection{Quantum well}
\label{subsec:qw}

Via L\"owdin perturbation theory (or the folding down approach), we obtain the following effective 2$\times$2 Hamiltonian for wurtzite wells
\begin{equation}
\label{eq:well3}
\mathcal{H}^{\rm 3D}=\mathcal {H}_{qw}+\mathcal{H}_{so}^R +\mathcal{H}_{so}^D, 
\end{equation}
where the first contribution is spin-independent and the last two terms correspond to the Rashba and Dresselhaus SO interactions, respectively. The spin independent part is given by
\begin{equation}
\label{eq:hqwcc}
   {\mathcal H}_{qw}(z)= -\dfrac{\hbar^2}{2}\dfrac{d}{d z}\dfrac{1}{ m_{\bot}(z)}\dfrac{d}{d z}+\dfrac{\hbar^2k^2_{\|}}{2m_{\|}(z)}+{V}_{\rm eff}(z)+ V_{e}(z),
\end{equation}
with $m_{\bot}$ and $m_{\|}$ the longitudinal and transversal effective masses, respectively, 
\begin{equation}
\label{eq:mzcc}
   \dfrac{1}{m_{\bot}(z)}=\dfrac{1}{m_0}+\dfrac{2P^2_1}{\hbar^2}\left[\dfrac{1}{ E_g+\Delta_1+\Delta_2}-\dfrac{ V_{e}(z)- V_C(z)}{( E_g+\Delta_1+\Delta_2)^2}\right], 
\end{equation}
\begin{eqnarray}
\label{eq:mxycc}
\nonumber
\dfrac{1}{ m_{\|}(z)}=\dfrac{1}{m_0}&+&\dfrac{P^2_2}{\hbar^2}  \left[\dfrac{1}{E_g}+\cfrac{1}{ E_g+2\Delta_2}\right. \\ 
&-&\left. \dfrac{V_{e}(z)-V_A(z)}{E^2_g}  - \dfrac{V_{e}(z)- {V}_B(z)}{(E_g+2\Delta_2)^2}\right].
\end{eqnarray}
The extra effective potential $V_{\rm eff}= V_D+V_s$ includes the Darwin term $V_D$ and an $s$--$p_z$ mixing-induced contribution $V_s$. The corresponding expressions read
\begin{equation}
\label{eq:vdcc}
   V_D(z)=\dfrac{P^2_1}{2(E_g+\Delta_1+\Delta_2)^2}\dfrac{d^2 V_{e}(z)}{dz^2},
\end{equation}
\begin{equation}
\label{eq:vscc}
    V_s(z)=\dfrac{2\Delta^2_{\rm sz}}{E_g+2\Delta_2}-\dfrac{2\Delta^2_{\rm sz}\left[V_{e}(z)- V_B(z)\right]}{(E_g+2\Delta_2)^2}.
\end{equation}

The spin-dependent Rashba Hamiltonian is given by 
\begin{equation}
  \label{eq:hsowell}
   \mathcal{H}_{so}^R =  \eta(z)\left(\sigma_x k_y - \sigma_y k_x\right),
  \end{equation}
with SO coupling parameter
\begin{equation}
   \label{eq:etadia}
    \eta(z) =\eta_{c}(z) +  \eta_{H}\dfrac{d V_{H}(z)}{dz} +\eta_{w}\dfrac{d V_{w-e}(z)}{dz}.
\end{equation}
In the expression above, each term corresponds to the bulk, Hartree, and structural contributions, respectively.

The bulk Rashba contribution, due to both the well and barrier layers, reads
\begin{eqnarray}
  \label{eq:etaccc}
   \eta_{c}(z)&=&\cfrac{2P_2\Delta_{\rm sz}}{ E_g+2\Delta_2}\left[1+\dfrac{4\Delta^2_3}{(E_g+\Delta_2)(E_g+\Delta_1+\Delta_2)}\right. \nonumber \\ 
   &-& \left.\dfrac{8\Delta_{\rm sz}^2}{(E_g+2\Delta_2)^2}-\dfrac{ V_{e}(z)- V_B(z)}{ E_g+2\Delta_2}\right].
\end{eqnarray}
Note that $\eta_c(z)$ takes different values in each layer, i.e., $\eta_c(|z|<L_w/2)\equiv\alpha_{\rm bulk}\rm{(well)}$ [see Eq.~(\ref{eq:rashbab})] and $\eta_c(|z|>L_w/2)\equiv\alpha_{\rm bulk}\rm{(barrier)}$~\cite{footnote-brashba}.~We should emphasize that there is no analogue of such a contribution to the SO interaction in zincblende systems, in which the $s$--$p_z$ mixing is not allowed by symmetry.

The Hartree and structural terms contribute only in third-order in the energy denominator,
\begin{equation}
  \label{eq:etahcc}
  \eta_{H}=\cfrac{P_1P_2\Delta_{3}}{( E_g+2\Delta_2)( E_g+\Delta_1+\Delta_2)}
  \left(\cfrac{1}{E_g+2\Delta_2}+\cfrac{1}{ E_g+\Delta_1+\Delta_2}\right) \\
\end{equation}
and \begin{equation}
  \label{eq:etawcc}
  \eta_{w}=-\dfrac{P_1P_2\Delta_{3}}{( E_g+2\Delta_2)( E_g+\Delta_1+\Delta_2)}
  \left(\dfrac{ \delta_C/ \delta_e}{ E_g+2\Delta_2}+\dfrac{  \delta_B/ \delta_e}{ E_g+\Delta_1+\Delta_2}\right). \\
\end{equation}

The Dresselhaus term, which in this case has the same symmetry as the Rashba, reads
\begin{eqnarray}
\label{eq:dreq1}
\mathcal{H}_{so}^D=\gamma\left(bk_z^2-k_{\|}^2\right)\left(\sigma_x k_y - \sigma_y k_x\right),
\end{eqnarray}
with $\gamma$ and $b$ determined in terms of bulk quantities, for details see Appendix \ref{app:hvq}.

The expressions shown above for the Rashba coupling do not account for possible \kp terms within the valence band subspace. These are rigorously nonzero by point-group symmetry arguments~\cite{Dugdale,Dargys}. It can be shown, however, that their contribution is negligible. In Appendix~\ref{app:hvq}, we obtain the corresponding expression of $H_v$ (valence band subspace) including the \kp interaction within the valence bands, charaterized by the parameter $Q=-(i\hbar/m_0)\bra{Z^\prime}p_x\ket{X}=(i\hbar/m_0)\bra{Z^\prime}p_y\ket{Y}$. The additional terms in the $2\times2$ electron Hamiltonian are also derived, and we confirm their negligible effect on the Rashba SO coupling. This \kp interaction contributes to the Dresselhaus SO coupling as well. Though, we emphasize that in our simulations the Dresselhaus interaction is put by hand by treating the bulk Dresselhaus coefficients as semi-empirical parameters.

\subsection{General expression}
\label{subsec:general}

For completeness, we present below the general effective Schr\"odinger equation for electrons in wurtzite heterostructures with an arbitrary confining potential and external magnetic field. More especifically, our general result is valid for quantum wells, wires, and dots [cf. Eq. (6.26) in Ref.~\cite{Winkler} for zincblende heterostructures].

The total $2\times2$ Hamiltonian reads
\begin{eqnarray}
\label{eq:heff3}
 H_{\rm eff}= H_0 +  \bm{\sigma} \cdot \bm{B}_{\rm eff}(\bm{r}) + H_{\rm so},
\end{eqnarray}
in which $H_0$ is spin-independent, $\bm{B}_{\rm eff}(\bm{r})$ is an effective magnetic field, and $H_{\rm so}$ is the SO coupling. The expression for $H_0$ is given by
\begin{widetext}
\begin{eqnarray}
\label{eq:heff30}
\nonumber H_{0}=\frac{\hbar^2}{2}\left[\bm{k}_{\|}\frac{1}{m_{\|}(\bm{r})}\bm{k}_{\|}+\bm{k}_z\frac{1}{m_{\bot}(\bm{r})}\bm{k}_z\right]+V_e(\bm{r})
&+&\frac{2\Delta^2_{sz}}{E_g+2\Delta_2}\left[1-\frac{V_e(\bm{r})-V_B(\bm{r})}{ E_g+2\Delta_2}\right]\nonumber \\
&+&\dfrac{P^2_1}{( E_g+\Delta_1+\Delta_2)^2}\dfrac{\partial^2 V_e(\bm{r})}{\partial z^2} 
+\left[\frac{P^2_2}{ E^2_g}+\frac{P^2_2}{( E_g+2\Delta_2)^2}\right]\nabla^2_{\|}V_e(\bm{r}), 
\end{eqnarray}
\end{widetext}
where the out-of-plane and in-plane effective masses, which now that depend on $\bm{r}$, are given by Eqs.~\eqref{eq:mzcc} and \eqref{eq:mxycc}, respectively. The last two terms in \eqref{eq:heff30} are equivalent to the Darwin term in the Pauli equation~\cite{Sakurai}.

The effective magnetic field can be written as
\begin{widetext}
 \begin{eqnarray}
 \label{eq:heff3b}
 \bm{B}_{\rm eff}(\bm{r})&=&-\dfrac{e}{\hbar}\dfrac{P_1P_2\Delta_3}{( E_g+\Delta_2)( E_g+\Delta_1+\Delta_2)}\left(1-\dfrac{V_e(\bm{r})-V_B(\bm{r})}{ E_g+2\Delta_2}
+ \dfrac{V_e(\bm{r})-V_C(\bm{r})}{ E_g+\Delta_1+\Delta_2}\right)(B_x\hat{\bm{x}} + B_y\hat{\bm{x}})
+\dfrac{e}{\hbar}\dfrac{P^2_2}{2}\left(-\frac{V_A(\bm{r})}{ E^2_g}+\frac{V_B(\bm{r})}{( E_g+2\Delta_2)^2}\right)B_z \hat{\bm{z}},  \nonumber \\
 \end{eqnarray}
\end{widetext} 
where $\bm{B}=\sum_{i=x,y,z}B_i\bm{\hat{i}}$ is the applied magnetic field. The SO Hamiltonian reads
\begin{widetext}
\begin{eqnarray}
 \label{eq:heff3so}
\nonumber H_{so}&=&\left(\frac{2P_2\Delta_{\rm sz} }{ E_g+2\Delta_2}+\frac{4P_2\Delta^2_3\Delta_{\rm sz} }{( E_g+2\Delta_2)^2( E_g+\Delta_1+\Delta_2)} -\frac{8P_2\Delta^3_{sz} }{( E_g+2\Delta_2)^3} \right) (\sigma_x k_y-\sigma_y k_x)
-\frac{P_2\Delta_{\rm sz}}{( E_g+2\Delta_2)^2}\left\{\sigma_xk_y-\sigma_yk_x,V_e(\bm{r})-V_B(\bm{r})\right \}\\\nonumber
&+&\frac{P^2_2}{2 E^2_g}\left[\bm{k}\times \bm{\nabla}V_A(\bm{r})\right]_z\sigma_z-\frac{P^2_2}{2( E_g+\Delta_2)^2}\left[\bm{k}\times \bm{\nabla}V_B(\bm{r})\right]_z\sigma_z \nonumber \\
&+&\frac{P_1P_2\Delta_3}{( E_g+2\Delta_2)^2( E_g+\Delta_1+\Delta_2)}\bm{\sigma}_{\|}\cdot
\left[\bm{k}\times \bm{\nabla}V_B(\bm{r})\right]_{\|}
+\frac{P_1P_2\Delta_3}{( E_g+2\Delta_2)( E_g+\Delta_1+\Delta_2)^2}\bm{\sigma}_{\|}\cdot
\left[\bm{k}\times \bm{\nabla}V_C(\bm{r})\right]_{\|} \nonumber \\
&+&\frac{P_2\Delta_{\rm sz}}{( E_g+2\Delta_2)^3}\left\{\sigma_xk_y-\sigma_yk_x,V^2_e(\bm{r})+V^2_B(\bm{r})\right\}
-\frac{P_2\Delta_{\rm sz}}{( E_g+2\Delta_2)^3}\left[\left\{V_e(\bm{r}), (\sigma_xk_y-\sigma_yk_x)V_B(\bm{r}) \right\}+\left\{V_B(\bm{r}), (\sigma_xk_y-\sigma_yk_x)V_e(\bm{r}) \right\}\right], \nonumber \\
\end{eqnarray}
\end{widetext}
with $\bm{k}$ the kinetic momentum, i.e., $\hbar\bm{k}=-i\hbar \nabla$+$e \bm{A}$, $\bm{B}=\nabla\times\bm{A}$ (real external field), which should be distinguished from the canonical momentum $\hbar\bm{k}=-i\hbar \nabla$.
The bracket $\{\cdot, \cdot\}$ stands for the anti-commutator. 

The general Hamiltonian in Eq.~\eqref{eq:heff3} is one of the main results of the present paper. We must point out a major difference between the latter and the effective model for zincblende structures~\cite{Winkler}: the nonequivalence between the $z$ direction with respect to $x$ and $y$ allows for the $s$--$p_z$ hybridization, which then leads to the emergence of additional terms (see coefficients $\propto \Delta_{\rm sz}$).

From the expressions above, it is quite simple to see this nonequivalence between $z$ and $x$, $y$. For the sake of simplicity, let us focus on the bulk case. In addition to having different effective masses in the longitudinal and transversal directions [see Eqs.~\eqref{eq:effmassz} and \eqref{eq:effmassp}, respectively], Eq.~\eqref{eq:heff3b} reveals an anisotropy in the effective $g$-factor, there is no $z$-component. This feature, as well as the analysis of the extra terms mentioned above, will be addressed elsewhere.

Note that, apart from the Dresselhaus term (not included in the derivation above), Eq.~\eqref{eq:well3} is a particular case of \eqref{eq:heff3}, see Eqs.~\eqref{eq:hqwcc}--\eqref{eq:etawcc}.

Our general Hamiltonian can be used  to study a variety of heterostructures. Next we focus on one- and two-subband quantum wells, for which we can write down an effective Hamiltonian and self-consistently determine all the relevant SO couplings.

\section{Effective $4\times 4$ Hamiltonian}
\label{sec:h2d}

From the 3D Hamiltonian of Sec.~\ref{subsec:qw}, we can derive an effective low-energy quasi-2D model. More specifically, here we obtain a model Hamiltonian for quantum wells with one and two subbands.

Our approach consists first of self-consistently determining the spin-degenerate eigenvalues $\mathcal{E}_{{\bm k_{\|}}\nu}$=$\mathcal{E}_\nu$ +$\hbar^2k^2_{\|}/2m_{\|}$ and corresponding eigenspinors $| {\bm k_{\|}}\nu\sigma\rangle=| {\bm k_{\|}}\nu\rangle \otimes |\sigma\rangle$, $\langle \bm{r}| {\bm k_{\|}}\nu\rangle=e^{i{\bm k_{\|}\cdot \bm r_{\|}}}\psi_\nu(z)$ of $H_{\rm qw}~\eqref{eq:hqwcc}$. Here $\mathcal{E}_\nu$ ($\psi_\nu$) is the $\nu^{\rm th}$ subband (wave function) of the well and $\sigma= \uparrow,\downarrow$ is the electron spin component along the $z$ direction. Then we project the total 3D Hamiltonian~\eqref{eq:well3} onto the basis  set $\{| {\bm k_{\|}}\nu\sigma\rangle \}$. 

Explicitly, the effective model with two subbands $\left\{|\bm{k}_{\|}1\uparrow\rangle, |\bm{k}_{\|}1\downarrow \rangle, |\bm{k}_{\|}2\uparrow \rangle, |\bm{k}_{\|}2\downarrow \rangle\right\}$, in the coordinate system ${x} || (100)$, ${y}||(010)$, can be written as
\begin{widetext}
\begin{eqnarray}
\label{eq:qw2d}
\mathcal{H}^{\rm 2D} =\left(
\begin{array}{cccc}
\mathcal{E}_{{\bm k_{\|}}1} & i\left(\alpha_1 +\beta_1\right)k_- & 0 & i\left(\eta +\Gamma\right)k_- \\
-i\left(\alpha_1 +\beta_1\right)k_+ & \mathcal{E}_{{\bm k_{\|}}1} & -i\left(\eta +\Gamma\right)k_+ & 0 \\
0 & i\left(\eta +\Gamma\right)k_- & \mathcal{E}_{{\bm k_{\|}}2} & i\left(\alpha_2 +\beta_2\right)k_- \\
-i\left(\eta +\Gamma\right)k_+ & 0 & -i\left(\alpha_2 +\beta_2\right)k_+ & \mathcal{E}_{{\bm k_{\|}}2}  
\end{array}
\right). 
\end{eqnarray}
\end{widetext}
Here we have neglected the $z$-dependence of $m_{\|}$ and chosen its value in the well layer since the electrons are mostly confined there. In Sec.~\ref{sec:hetero}, we evaluate $m_{\|}$ and show it is in great agreement with experimental results. 

The two $2\times2$ blocks [upper left ($\alpha_1$, $\beta_1$) and lower right ($\alpha_2$, $\beta_2$)] of Eq.~\eqref{eq:qw2d} correspond to the usual Rashba-Dresselhaus Hamiltonian coupled by the ``off-diagonal'' intersubband block ($\eta$, $\Gamma$). Note, however, that differently from zincblende materials, the Rashba and Dresselhaus terms have the same symmetry.

The intrasubband (intersubband) Rashba $\alpha_\nu$ ($\eta)$ and Dresselhaus $\beta_\nu$ ($\Gamma$) SO coefficients are defined as follows
  \begin{eqnarray}
 \eta_{\nu \nu^\prime} = \langle \psi_\nu |  \eta_c(z)+\eta_H \partial_z  V_H (z) +\eta_w \partial_z V_{w-e}(z) |\psi_{\nu^\prime} \rangle,
\label{eq:eta}
\end{eqnarray}
and
\begin{eqnarray}
 \Gamma_{\nu \nu^\prime}=\gamma \big(b\langle \psi_\nu |k^2_z|\psi_{\nu^\prime} \rangle-k_{\|}^2\delta_{\nu\nu^\prime}\big),
\label{eq:gamma}
\end{eqnarray}
where the Rashba coefficients are $\alpha_\nu \equiv\eta_{\nu \nu}$, within each subband, and $ \eta \equiv \eta_{12}$, between subbands. The same is valid for the Dresselhaus coefficients $\beta_{\nu} \equiv \Gamma_{\nu \nu}$ and $\Gamma \equiv \Gamma_{12}$.

The Rashba coupling $\alpha_\nu$ ($\eta$) can be written in terms of  several contributions, i.e., $\alpha_\nu=\alpha_\nu^{\rm c}+\alpha_\nu^{\rm H}+\alpha_\nu^{\rm w}$, being $\alpha_\nu^{\rm c}=\langle \psi_\nu | \eta_c(z)|\psi_\nu \rangle$ the bulk coefficient~\cite{footnote-bulk}, $\alpha_\nu^{\rm H}=\eta_{\rm H} \langle \psi_\nu | \partial_z {V}_{\rm H}(z)|\psi_\nu \rangle$ the Hartree term, and  $\alpha_\nu^{\rm w}=\eta_{\rm w} \langle \psi_\nu | \partial_z  {V}_{w-e}(z)|\psi_\nu \rangle$ the contribution due to the structural potential. The Hartree coefficient can also be split into different contributions [see Eq.~\eqref{eq:VHartree}],
\begin{equation}
\alpha_{\nu}^{\rm H} = \alpha_{\nu}^{\rm elect} + \alpha_{\nu}^{\rm d} +\alpha_{\nu}^{\rm int} + \alpha_{\nu}^{\rm g},
\end{equation}
where
\begin{equation}
\label{eq:aelect}
\alpha_\nu^{\rm elect} = \eta_{\rm H} \langle \psi_\nu | \partial_z {V}_{\rm elect}(z)|\psi_\nu \rangle,
\end{equation}
\begin{equation}
\label{eq:ad}
\alpha_\nu^{\rm d} = \eta_{\rm H} \langle \psi_\nu | \partial_z {V}_{\rm d}(z)|\psi_\nu \rangle,
\end{equation}
\begin{equation}
\label{eq:aint}
\alpha_\nu^{\rm int} = \eta_{\rm H} \langle \psi_\nu | \partial_z {V}_{\rm int}(z)|\psi_\nu \rangle,
\end{equation}
and
\begin{equation}
\label{eq:ag}
\alpha_\nu^{\rm g} = \eta_{\rm H} \langle \psi_\nu | \partial_z {V}_{\rm g}(z)|\psi_\nu \rangle
\end{equation}
are due to, respectively, the purely electron Hartree potential, the doping potential, the internal potential (built-in electric fields), and the external gate potential. Note that all these SO coupling coefficients depend on the self-consistent potential and the subband wave functions.

Interestingly, as mentioned above, the Rashba and Dresselhaus terms induce the same electron spin configuration in a $c$ axis oriented wurtzite structure~\cite{gloub:2014,wu:2010}, in contrast to the (001)-grown zincblende heterostructures. This allows us to define an effective SO coupling parameter for each subband $\alpha_{\nu}^{\rm eff}=\alpha_\nu+\beta_{\nu}$, and similarly an effective intersubband SO term $\eta^{\rm eff}=\eta+\Gamma$. Therefore, experimental measurements of the SO splittings correspond to the effective coupling coefficients $\alpha_{\nu}^{\rm eff}$ and $\eta^{\rm eff}$.
In the following, we evaluate these quantities by self-consistently solving Schr\"odinger and Poisson equations.

\section{Results and discussions}
\label{sec:algan}

Here we first introduce the structure of the wells. Then we discuss our calculated SO couplings based on the model presented in the previous sections. Our discussions cover both the single and double well cases with either one or two occupied subbands.

\subsection{Heterostructures}
\label{sec:hetero}
We consider GaN/Al$_{x}$Ga$_{1-x}$N heterostructures grown along the $z||$[0001] direction.
For the single-well, our heterostructure is defined by a well of width $L_{w}=10$~nm and two symmetric barrier regions of width $L_{b}=7$~nm [see Fig.~\ref{fig1}(b)].
Unless otherwise stated, the electron density is \textit{fixed} at $n_e=1.0\times10^{12}$ cm$^{-2}$, arising from two symmetrically doped layers sitting $6$~nm away from the center of the well.
Our double well has a similar geometry, except for an additional barrier of width $L_{cb}=2$~nm embedded in the center of the structure.

Wurtzite GaN-based heterostructures are usually grown on a GaN substrate by molecular beam epitaxy, i.e., the  Al$_{x}$Ga$_{1-x}$N barriers are deformed and the corresponding 
lattice constant is adjusted to the GaN substrate and quantum wells~\cite{Grandjean,Pokatilov,Park1}, which is the case we consider here.
In the Al$_x$Ga$_{1-x}$N layers~\cite{Vurgaftman}, we choose the SO and the crystal field splitting parameters $\Delta_{\rm so}=0.014$~eV, $\Delta_{\rm cr}=0.019-0.183x$~eV, respectively, the lattice constant $A_l=3.189-0.077x$~\AA, the deformation potentials $D_1=-3.0$~eV, $D_2=3.6$~eV, $D_3=8.82+0.78x$~eV, $D_4=-4.41+0.39x$~ eV (valence band),
 $a_{c1}=-9.5-2.5x$~eV, $a_{c2}=-8.2+2.8x$~eV (conduction band)~\cite{Shan}, and the elastic constants $C_{13}=106+2x$~Gpa, $C_{33}=398-25x$~ Gpa~\cite{Polian}.
 The unstrained band gap is $E_g=3.507+2.723x$~eV and the strained one is calculated by taking into account the strain-induced band edge shifts [see Sec.~\ref{sec:ccs}].
The spontaneous polarization is given by $-0.029-0.051x$~C$\cdot$m$^{-2}$ and the piezoelectric coefficients by $e_{13}=-0.35-0.15x$~C$\cdot$m$^{-2}$, $e_{33}=1.27+0.25x$~C$\cdot$m$^{-2}$~\cite{Vurgaftman}.

\begin{table}[h]
	\caption{Band parameters (in eV) for GaN/Al$_{x}$Ga$_{1-x}$N heterostructures, see the single well shown in Fig.~\ref{fig1}(b).}
	\begin{ruledtabular}
		\begin{tabular}{ccccc}
			$x$ & $ \delta_e(\delta_{be})$ & $\delta_9(\delta_{bA})$ & $\delta_B(\delta_{bB})$ & $\delta_{C}(\delta_{bC})$ \\ \hline
			
			$0.30$ & $0.5252$ & $0.2950$ & $0.1546$ & $0.2784$\\
			$0.40$ & $0.7051$ & $0.4044$ & $0.2098$ & $0.3877$\\
			$0.50$ & $0.8876$ & $0.5157$ & $0.2059$ & $0.4990$\\
			$0.60$ & $1.0728$ & $0.6288$ & $0.3228$ & $0.6120$\\
			$0.70$ & $1.2608$ & $0.7438$ & $0.3805$ & $0.7270$\\
			$0.80$ & $1.4515$ & $0.8606$ & $0.4391$ & $0.8438$\\
		\end{tabular}
	\end{ruledtabular}
	\label{table:bandparameter}
\end{table}

The fraction of the $s$--$p_z$ mixing we consider here is of $\sim1\%$. This is the same value used for ZnO, which has a very similar band gap. We then obtain $\Delta_{\rm sz}$ $\sim$ $0.467$~meV~\cite{Voon,mixing}. The interband Kane parameters are taken as $E_{1}$=$E_2$=14.0 eV~\cite{Yan,Vurgaftman,Ambacher1,Im1}, with $E_{1,2}$=${2m_0}P^2_{1,2}/{\hbar^2}$, and are assumed the same in the well and barriers. We choose a uniform dielectric constant $\epsilon_r$=10.0~\cite{Park} and consider the temperature $T=2$~K.
The relevant band parameters for GaN/Al$_{x}$Ga$_{1-x}$N heterostructrures are shown in Table~\ref{table:bandparameter} for typical Al contents on the barriers [see
Fig.~\ref{fig1}(b)].

By inserting the relevant parameters listed above into Eqs.~\eqref{eq:mzcc} and \eqref{eq:mxycc}, we self-consistently determine the out-of-plane $m_{\bot}$(GaN)$=0.2014m_0$ and in-plane $m_{\|}$(GaN)=0.2005$m_0$ effective masses, in excellent agreement with the experimental values, 0.2$m_0$ and 0.2$m_0$~\cite{Vurgaftman,Barker}, respectively.
We also calculate the Rashba SO constants $\alpha_{\rm bulk}$(GaN)=1.938 meV$\cdot$\AA, $\eta_{H}$=0.01138~\AA$^2$, and $\eta_{w}$=2.462~meV$\cdot$ \AA$^2$, see Eqs. \eqref{eq:etaccc}--\eqref{eq:etawcc}. For the Dresselhaus parameters $\gamma$ and $b$ [see Eq.~\eqref{eq:dreq1}],  we use $\gamma=0.32$~eV$\cdot$\AA$^3$ and $b=3.855$~\cite{Fu,Wang}.

\subsection{Single well}
\label{sec:sw}
Next we show our calculated SO coupling coefficients for a GaN/Al$_{0.3}$Ga$_{0.6}$N single well with one occupied subband. The dependence of these SO couplings on the Al content $x$ on the Al$_{x}$Ga$_{1-x}$N layers is discussed as well. We also determine the SO couplings of a GaN/Al$_{0.3}$Ga$_{0.6}$N well with two occupied subbands. 

\begin{figure}[h]
	\includegraphics[width=8.0cm] {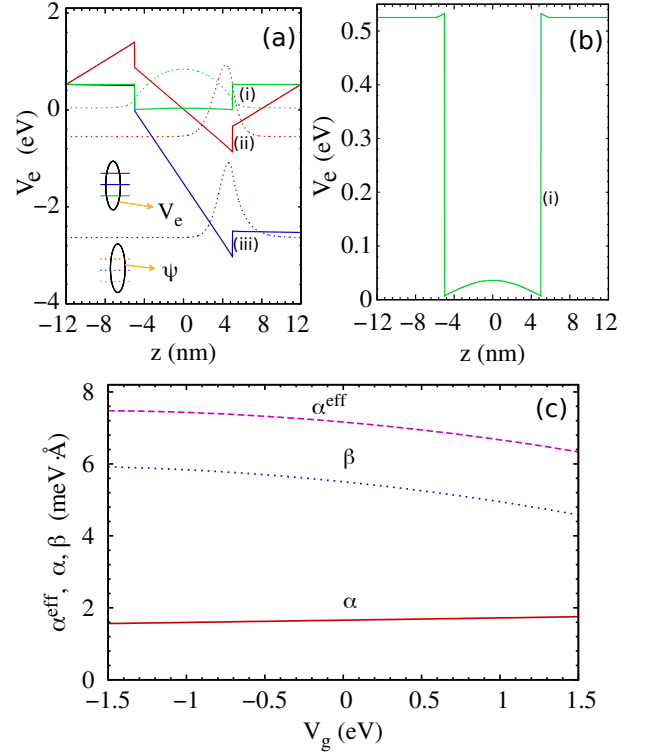}
	\caption{(Color online) (a) Self-consistent potential $V_{e}$ (solid curves) and wave function profile $\psi$ of the first subband (dotted curves) for three different conditions of the built-in fields, for a GaN/Al$_{0.3}$Ga$_{0.7}$N single well at $V_{\rm g}=0$. (i) green curves: zero built-in field (flat-band model); (ii) red curves: built-in field with periodic boundary conditions; (iii) blue curves: built-in field with neutral surface charge boundary conditions. The energy level (not shown) of the first subband for these three conditions are $\mathcal{E}_1=44.9, -551.2, -2634.2$ meV, respectively. (b) Blowup of the potential $V_e$ in the flat-band model. (c) Rashba $\alpha$, Dresselhaus $\beta$, and effective SO $\alpha^{\rm eff}=\alpha+\beta$ couplings as functions of $V_{\rm g}$. The areal electron density is kept fixed at $1.0\times 10^{12}$cm$^{-2}$, so that only the first subband is occupied. The temperature is $T=2$ K.
	}
	\label{fig3}
\end{figure}

\subsubsection{One occupied subband}
\label{sec:onebandsw}
Before discussing the SO couplings in detail, let us first have a look at our self-consistent solutions. Figure~\ref{fig3}(a) shows the profiles of the conduction band potential $V_{e}$ and of the first subband wave function of a GaN/Al$_{0.3}$Ga$_{0.7}$N single well with only one occupied subband. We consider three cases: (i) the flat-band model (no built-in field) (green curves), (ii) periodic boundary conditions (red curves), and (iii) neutral surface charge boundary conditions (blue curves).  

We observe that the flat-band model produces the usual profile of a confining square well potential [see Fig.~\ref{fig3}(b)] and its corresponding envelope wave function.
In contrast, in the presence of the strong built-in field (spontaneous and piezoelectric, $\sim$~MV/cm~\cite{Ambacher,Ambacher1}), both the periodic and the neutral surface charge boundary conditions
transform the rectangular-type well into a triangular-like one. Accordingly, electrons are mainly confined near one of the well/barrier interfaces.  We find that the SO couplings calculated in GaN wells in both cases are comparable, the discrepancy being of just $\sim0.2$~meV$\cdot$\AA.
Thus in the following, we focus on the widely adopted periodic boundary conditions~\cite{Park3,Ng}.
\begin{figure}[bth!]
\includegraphics[width=8.0cm] {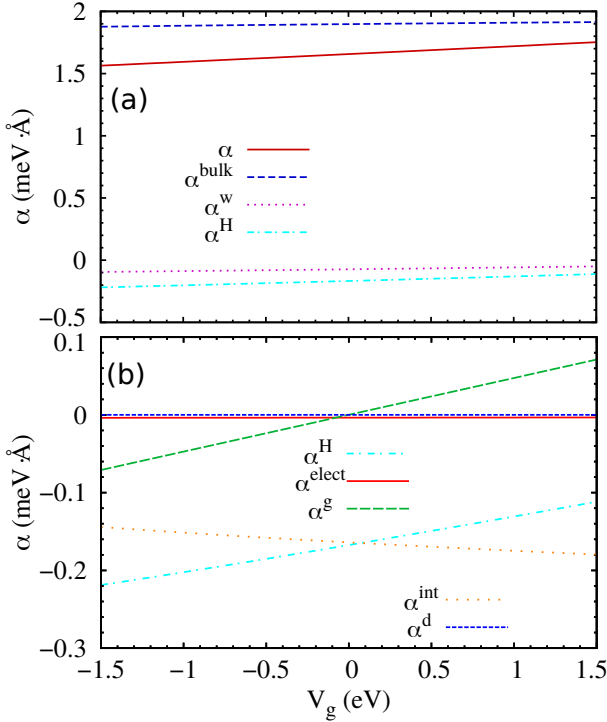}
\caption{(Color online) (a) Rashba $\alpha$ and its several contributions: the bulk $\alpha_{\rm bulk}$, the structural $\alpha^{\rm w}$, and the Hartree $\alpha^{\rm H}$ contributions as functions of $V_{\rm g}$ in a GaN/Al$_{0.3}$Ga$_{0.7}$N single well with one occupied subband.  
(b) Contributions to $\alpha^{\rm H}$: the pure electron Hartree  $\alpha^{\rm elect}$, the internal field $\alpha^{\rm int}$, the doping $\alpha^{\rm d}$, and the external gate $\alpha^{\rm g}$ coefficients as functions of $V_{\rm g}$.  
The areal electron density is held fixed at $1.0\times 10^{12}$cm$^{-2}$. The temperature is $T=2$ K.
}
\label{fig4}
\end{figure}

Figure~\ref{fig3}(c) shows the Rashba $\alpha$, Dresselhaus $\beta$, and effective SO $\alpha^{\rm eff}=\alpha+\beta$ couplings as functions of the gate voltage $V_{\rm g}$.
One can see that the Rashba term is relatively weaker than the Dresselhaus coupling, but their magnitudes are comparable.
At zero bias $V_{\rm g}=0$, we obtain the total SO coupling intensity $\alpha^{\rm eff}=7.16$ meV$\cdot${\AA}, in agreement with results from weak antilocalization measurements, in which the spin splitting parameter was reported ranging from 5.5 to 10.01~meV$\cdot${\AA}~\cite{Belyaev,Kurdak,Schmult,Thillosen1,Thillosen2}.
In addition, we find that the Rashba coupling, which in general strongly depends on $V_{\rm g}$ in usual zincblende quantum wells, remains essentially constant as $V_{\rm g}$ varies.
The weak dependence of $\alpha$ on $V_{\rm g}$ follows from the interplay of its several different contributions, as we discuss next.

In Fig.~\ref{fig4}(a) we show each contribution of $\alpha$ as a function of $V_{\rm g}$. Figure~\ref{fig4}(b) further shows the several terms of $\alpha^{\rm H}$ [see Eqs.~\eqref{eq:aelect}--\eqref{eq:ag}]. For the GaN and Al$_{0.3}$Ga$_{0.7}$N layers, we obtain $\alpha_{\rm bulk}\rm{(GaN)}=1.938$ meV$\cdot$\AA~ and $\alpha_{\rm bulk}\rm{(AlGaN)}=1.570$ meV$\cdot$\AA. Since the wave function is mostly confined in the well (GaN layer), the intrinsic bulk contribution $\alpha_{\rm bulk}\approx \alpha_{\rm bulk}\rm{(GaN)}$ remains essentially constant as $V_{\rm g}$ varies.

The combined contributions of the Hartree $\alpha^{\rm H}$ and structural $\alpha^{\rm w}$ terms refer to the usual Rashba coupling induced by the structural inversion asymmetry of the system. As can be seen in Figs.~\ref{fig4}(a) and~\ref{fig4}(b), the usual Rashba term is approximately one tenth of the bulk Rashba contribution. Therefore, the total Rashba coupling has a similar behavior with $V_g$ as the bulk Rashba term, i.e., it is weakly gate-dependent. In available reports on the SO coupling in wurtzite heterostructures concerning \kp interactions, however, the bulk Rashba term was missed~\cite{Chuang,Litvinov1,li:2011}.
On the other hand, we must emphasize that, even though the usual Rashba term is much weaker than the bulk Rashba coupling in GaN-based wurtzite wells, its contribution in relatively narrow gap semiconductors (e.g., GaAs in the wurtzite phase) can become important, thus possibly leading to a sensitive electrical control of it.   
\begin{figure}[bth!]
\includegraphics[width=8.0cm] {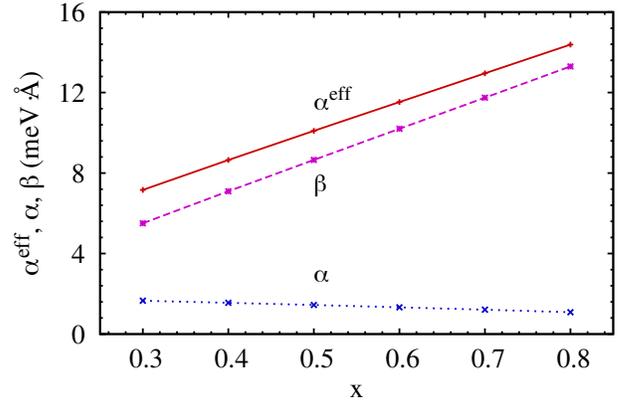}
\caption{(Color online) Rashba $\alpha$, Dresselhaus $\beta$, and the effective SO $\alpha^{\rm eff}=\alpha+\beta$ strengths in GaN/Al$_{x}$Ga$_{1-x}$N single wells with one occupied subband as functions of the Al content $x$ at $V_{\rm g}=0$. The areal electron density $n_e$ is held fixed at $1.0\times 10^{12}$cm$^{-2}$. The temperature is $T=2$ K.
}
\label{fig5}
\end{figure}

Let us now consider the effects on the SO couplings of the Al content on the Al$_{x}$Ga$_{1-x}$N layers. The Al $x$ content modifies the strength of the built-in field and the band offsets of the well, consequently changing the SO couplings.
Figure~\ref{fig5} shows $\alpha$, $\beta$, and $\alpha^{\rm eff}$ as functions of $x$ at $V_{\rm g}=0$. We find that the Dresselhaus coupling $\beta$ increases almost linearly  with $x$.
This is due to an increase of the piezoelectric polarization in the barriers as $x$ varies, which makes the electrons more confined in the well. On the other hand, we observe that the Rashba term $\alpha$ is not sensitive to $x$, again because of the dominant bulk contribution. 

\subsubsection{Two occupied subbands}
\label{sec:twobands}
Here we change the doping conditions so we can effectively occupy the second subband. In contrast to the case of one occupied subband, in which we have two symmetrically doped layers, we have now only a one-side doping layer (asymmetric doping), so that the doping field can partially compensate the built-in field, thus making the well confinement profile less steep. In addition,   
we increase the areal electron density to $n_e=1.5\times$10$^{13}$~cm$^{-2}$. This corresponds to a higher doping density $n_d$ in the doping layer. Since we assume $n_d=n_e$, we ensure charge neutrality in our system.

\begin{figure}[h]
	\includegraphics[width=8.0cm] {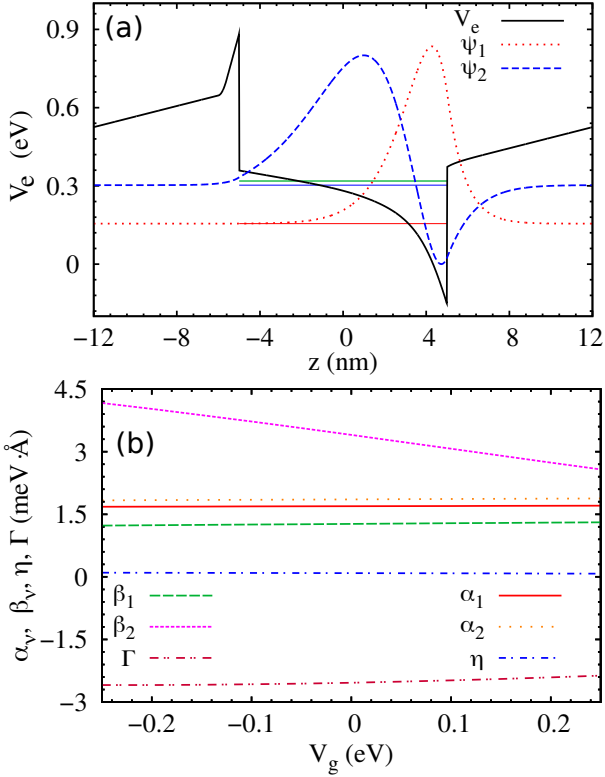}
	\caption{(Color online) (a) Self-consistent potential profile $V_{\rm e}$ and wave function profile of the first $\psi_1$ and second $\psi_2$ occupied subbands of a GaN/Al$_{0.3}$Ga$_{0.7}$N one-side doped single well at $V_{\rm g}=0$. The horizontal blue, red, and green lines inside the well indicate the subband energy levels $\mathcal{E}_1=155.1$ meV, $\mathcal{E}_2=302.8$ meV,  and the Fermi level $\mathcal{E}_{\rm F}=318.5$ meV, respectively.
		(b) Intrasubband (intersubband) Rashba $\alpha_\nu$ ($\eta$) and Dresselhaus $\beta_\nu$ ($\Gamma$) SO couplings as functions of $V_{\rm g}$. 
		The areal electron density is held fixed at $n_e=1.5\times 10^{13}$cm$^{-2}$. The temperature is $T=2$ K.
	}
	\label{fig6}
\end{figure}

Figure~\ref{fig6}(a) shows the self-consistent confining potential $V_{e}$ and the wave function profile of the first $\psi_1$ and second $\psi_2$ occupied subbands of the well at $V_{\rm g}=0$.  Because of the change in the doping conditions, we observe that the well becomes flatter [cf. Figs.~\ref{fig3}(a) and~\ref{fig6}(a)].
In Fig.~\ref{fig6}(b), we show the intrasubband  Rashba $\alpha_{\nu}$ and Dresselhaus $\beta_{\nu}$, as well as the intersubband Rashba $\eta$ and Dresselhaus $\Gamma$ SO couplings as functions of $V_{\rm g}$. The coupling $\alpha_\nu$ depends very weakly on $V_g$, similarly to the case of one occupied subband [cf. Figs.~\ref{fig3} and \ref{fig6}], because of the dominant contribution of the bulk Rashba term. The various constituents of $\alpha_\nu$ are not shown since they all have the same behavior as in the one occupied subband case.

As for the coupling $\beta_\nu$, as the confining potential becomes flatter, we find that the first subband term $\beta_1$ is around four times weaker than that in the well with only one subband occupied. In addition, $\beta_1$ remains almost constant as $V_{\rm g}$ varies. On the other hand, for the second subband, $\beta_2$ is relatively sensitive to $V_{\rm g}$. We attribute the different behaviors of $\beta_1$ and $\beta_2$ with $V_{\rm g}$, to the distinct distributions of electrons occupying the first and second subbands. Electrons occupying the first subband are mostly confined to the right well/barrier interface (i.e., a narrow triangular confinement) [see $\psi_1$  in Fig.~\ref{fig6}(a)], and cannot ``feel'' an overall modification of the potential due to $V_{\rm g}$. However, electrons occupying the second band spread almost all over the whole well region [see $\psi_2$ in Fig.~\ref{fig6}(a)]. Thus, the second subband wave function profile $\psi_2$ is more strongly dependent on $V_{\rm g}$, further leading $\beta_2=\gamma(b\langle\psi_2|k_z^2|\psi_2\rangle -k_{F,2}^2)$ to be more sensitive to $V_{\rm g}$ as compared to $\beta_1$ [see Fig.~\ref{fig6}(b)]. Here $k_{F,\nu}=\sqrt{2\pi n_{\nu}}$ is the Fermi wave vector of the $\nu$th subband and  $n_{\nu}$ its  occupation.

The intersubband Rashba coupling strength $\eta$ is much weaker than the intrasubband coupling $\alpha_\nu$. We attribute this to the fact that the bulk intersubband Rashba contribution $\eta_{\rm bulk}=\langle \psi_1|\eta_c(z)|\psi_2\rangle$ to $\eta$ is negligible, as electrons are mostly confined inside the well.  More specifically,  we have $\eta_{\rm bulk}=\langle \psi_1|\eta_{\rm c}(z)|\psi_2\rangle\approx\langle \psi_1|\alpha_{\rm bulk}({\rm GaN})|\psi_2\rangle \sim 0$.  Note that for $\alpha_\nu$, the bulk term $\alpha_{\rm bulk}$ dominates over the usual Rashba contributions.
On the other hand, for the intersubband Dresselhaus term $\Gamma=\langle \psi_1|k_z^2|\psi_2\rangle$, we find that it is comparable to $\beta_\nu$.  We should emphasize that $\Gamma$ vanishes in the symmetric configuration because of parity. Here $\Gamma$ is nonzero at $V_{\rm g}=0$, due to the presence of both the built-in field and the asymmetric one-side doping field. 

\subsection{Double well}
\label{sec:dw}
In this section, we focus on obtaining the SO couplings for the double quantum well GaN/Al$_{0.3}$Ga$_{0.7}$N with an Al$_{x}$Ga$_{1-x}$N central barrier (see Sec.~\ref{sec:hetero}).  

\begin{figure}[bth!]
\includegraphics[width=8.0cm] {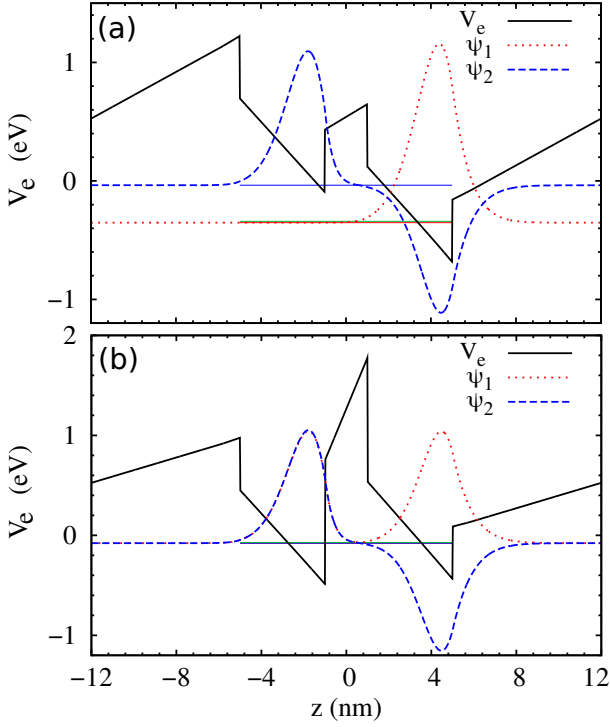}
\caption{(Color online) Self-consistent potential $V_{\rm e}$ and wave function profiles $\psi_\nu$ ($\nu=1,2$) of a GaN/Al$_{0.3}$Ga$_{0.7}$N double well with an Al$_{x}$Ga$_{1-x}$N central barrier at $V_{\rm g}=0$. (a) One occupied subband with $x=0.3$ in the central barrier; (b) two occupied subbands with $x=0.69$ in the central barrier. For comparison, in (a) it is shown the second subband wave function $\psi_2$ for the empty second level. The horizontal blue, red, and green lines inside the well indicate the subband energy levels $\mathcal{E}_1$, $\mathcal{E}_2$,  and the Fermi level $\mathcal{E}_{\rm F}$, respectively. In (a), $\mathcal{E}_1=-352.3$ meV,  $\mathcal{E}_2=-336.6$ meV, and $\mathcal{E}_{\rm F}=-340.4$ meV; in (b),
$\mathcal{E}_1=-78.2$ meV,  $\mathcal{E}_2=-78.2$ meV, and $\mathcal{E}_{\rm F}=-72.2$ meV.
The areal electron density $n_e$ is held fixed at $1.0\times 10^{12}$cm$^{-2}$.
}
\label{fig7}
\end{figure}

Before discussing the SO couplings in detail, let us first have a look at how the Al content in the central barrier affects the double well configuration. At low Al concentration, we find that the electrons are mostly confined to one side [right side of our wells, see Fig.~\ref{fig7}(a)] of the system, due to the presence of the strong built-in field, which makes the double well essentially an \emph{effective} single well. 
Figure~\ref{fig7}(a) shows the electron confining potential $V_{e}$ and the wave function profiles $\psi_\nu$ ($\nu=1,2$) of the lowest two subbands in a double well with $x=0.3$ in the central barrier.  Although only one subband is occupied in this configuration, for comparison, we also show $\psi_2$ for the empty second level of the well. As can be easily seen, this double-well configuration is essentially similar to the ones of the single wells discussed in previous sections.   

If we further increase the Al content in the central barrier, we find that the left and right sides of the double well compete to confine the electrons. Interestingly, when $x\sim 0.69$,  a \emph{seemingly} symmetric configuration occurs. In Fig.~\ref{fig7}(b), we show the corresponding confining potential and the wave function profiles at this configuration with two occupied subbands. We can see that the electrons occupying the two subbands are almost equally distributed on the left and right sides of the well, in contrast to the case of smaller values of $x$ [cf. Figs.~\ref{fig7}(a) and \ref{fig7}(b)]. However, we emphasize that this seemingly symmetric configuration (at $x=0.69$) is actually structurally asymmetric, since the gradient of the potential (i.e., the electric field) has the same sign (negative) on the left and right wells, see Fig.~\ref{fig7}(b). 
Next we determine the SO coupling coefficients for this configuration and discuss how these change with $V_{\rm g}$.  
        
\begin{figure}[bth!]
\includegraphics[width=8.0cm] {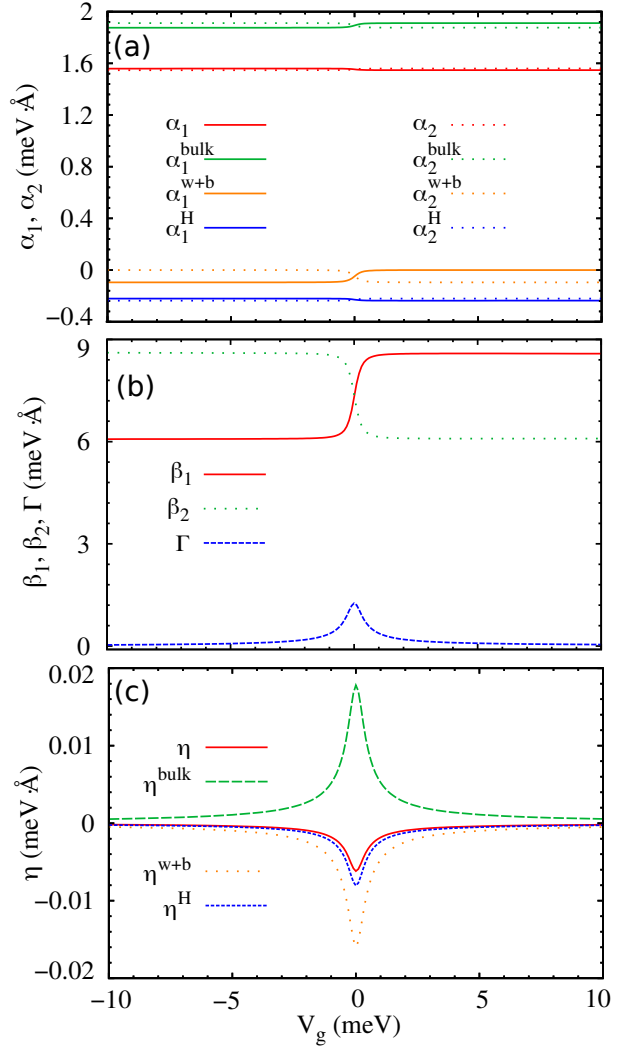}
\caption{(Color online) (a) Intrasubband Rashba couplings $\alpha_\nu$ ($\nu=1,2$) for a GaN/Al$_{0.3}$ Ga$_{0.7}$N double well with an Al$_{0.69}$Ga$_{0.31}$N central barrier as functions of $V_{\rm g}$. Several distinct constituents of $\alpha_\nu$: the bulk $\alpha_\nu^{bulk}$, the structural $\alpha_\nu^{w}$, and the total Hartree $\alpha_\nu^{H}$ contributions. (b) Intrasubband $\beta_\nu$ and intersubband $\Gamma$ Dresselhaus SO strengths as functions of $V_{\rm g}$. (c) Intersubband $\eta$ and its various contributions versus $V_{\rm g}$. The areal electron density is held fixed at $1.0\times10^{12}$ cm$^{-2}$.
}
\label{fig8}
\end{figure}

Figure~\ref{fig8}(a) shows $\alpha_\nu=\alpha_\nu^{\rm bulk}+\alpha_\nu^{\rm H}+\alpha_\nu^{\rm w+b}$ and their several different contributions: $\alpha_\nu^{\rm bulk}$, $\alpha_\nu^{\rm H}$, and $\alpha_\nu^{\rm w+b}$, with $\alpha_\nu^{\rm w+b}\equiv\alpha_\nu^{\rm w}+\alpha_\nu^{\rm b}$, for each subband $\nu=1,2$. Note that the structural term in a double well has an additional contribution $\alpha^{\rm b}_{\nu}=\eta_{\rm b}\expval{\frac{dV_b(z)}{dz}}{\nu}$ due to the presence of the central barrier~\cite{Esmerindo1}. The potential $V_b(z)$ describes the structural potential between the well and the central barrier. Straightforwardly, $\eta_b$ has the same expression as $\eta_w$ [see Eq.~\eqref{eq:etadia}] with the well offsets $\delta_i$ being replaced by the central barrier offsets $\delta_{bi}$. Similarly to the case of single wells, we find that $\alpha_\nu$ remains essentially constant as a function of $V_{\rm g}$, since the bulk contribution (not sensitive to $V_{\rm g}$) dominates over all the other contributions. Interestingly, we find that $\alpha_1$ and $\alpha_2$ almost interchange their respective values, across the seemingly symmetric configuration (at $V_{\rm g}=0$), because electrons occupying the first and second subbands interchange their distributions for $V_{\rm g}>0$ and $V_{\rm g}<0$.  This feature of interchanging values across $V_{\rm g}=0$ also holds for the Dresselhaus terms $\beta_1$ and $\beta_2$, with even more pronounced behavior (cf. $\alpha_\nu$'s), as shown in Fig.~\ref{fig8}(b).  

In Figs.~\ref{fig8}(b) and \ref{fig8}(c), we show the intersubband Dresselhaus $\Gamma$ and the Rashba $\eta$ couplings, respectively,  as functions of $V_{\rm g}$. We find that both $\eta$ and $\Gamma$ achieve their maximal strength at $V_{\rm g}=0$, in contrast to the intrasubannd $\alpha_\nu$ and $\beta_\nu$. This arises from the intersubband SO couplings depending on the overlap of the wave functions $\psi_1$ and $\psi_2$ of the two subbands, see Eqs.~\eqref{eq:eta} and \eqref{eq:gamma}. At $V_{\rm g}=0$ (seemingly symmetric configuration), $\psi_1$ and $\psi_2$ have a perfect overlap. However, away from $V_{\rm g}=0$, $\psi_1$ and $\psi_2$ are largely separated, i.e., one is mostly confined in the left well and the other in the right well.

On the other hand, for large Al concentration in the central barrier ($x\sim 0.84$), we find that the one subband electron occupation is restored (for the same areal electron density), similarly to the case of low $x$ mentioned previously, except that we have an almost reversed configuration of the left and right wells constituting the double well structure.

\section{Concluding remarks}
\label{sec:summary}

We have derived a very general SO Hamiltonian for conduction electrons in wurtzite heterostructures valid for arbitrary confining potentials (quantum wells, wires, and dots) and external magnetic fields. Our starting point is the $8\times8$ Kane model including 
the $s$--$p_z$ orbital mixing relevant to wurtzite systems. We then obtain a $2\times2$ effective SO Hamiltonian by applying both the Löwdin perturbation theory and the folding down approach. For concreteness, we focus on GaN/AlGaN wells and investigate in detail the electron SO couplings.

In addition to the $s$--$p_z$ orbital mixing (absent in zincblende structures), we have taken into account the renormalization of the conduction band spinor component when deriving the effective 2$\times$2 electron Hamiltonian for heterostructures; these two elements were not included in previous works. Most importantly, we find that these are crucial in obtaining the bulk Rashba and Dresselhaus terms (within the eight-band approach), and make the effective electron Hamiltonian energy independent (i.e., a \emph{real} Schr\"odinger-type equation). 
 
Through a self-consistent Schr\"odinger-Poisson calculation on GaN/AlGaN both single and double wells, involving either one or two occupied subbands, we have determined all the relevant SO strengths, i.e., the intrasubband Rashba $\alpha_\nu$ ($\nu=1,2$) and Dresselhaus $\beta_\nu$ couplings, as well as the intersubband Rashba $\eta$ and Dresselhaus $\Gamma$ couplings. Our calculated spin-orbit coupling is in agreement with experimental data. We have also determined the several distinct constituents of the SO couplings due to the bulk, the Hartree, and the structural contributions, and find that the SO couplings follow from the interplay of all these terms. 
Particularly, we find that the bulk Rashba term dominates over all other remaining contributions to the intrasubband Rashba couplings. For our double wells, we find a \emph{seemingly} symmetric configuration, at which both the intersubband Rashba and Dresselhaus couplings assume their largest values, due to a perfect overlap of the wave functions of the two subbands. On the other hand, across this configuration when we vary $V_{\rm g}$,  the intrasubband Dresselhaus couplings $\beta_1$ and $\beta_2$ almost interchange their values, a similar behavior also occuring for the Rashba couplings $\alpha_1$ and $\alpha_2$. In addition, we have derived an effective spin-orbit Hamiltonian for electrons with arbitrary confining potentials and external magnetic fields, valid for quantum wells, wires, and dots.   

As a final remark, although the SO couplings in our GaN wells are not sensitive (except for the double well around the seemingly symmetric configuration) to $V_{\rm g}$, we emphasize that a sensitive electrical control of the Rashba couplings could be possible in narrow-gap wurtzite wells (e.g., GaAs with wurtzite structure), where the usual Rashba contributions (sensitive to $V_{\rm g}$) could be comparable or even larger than the bulk contribution. Accordingly, since the Rashba and Dresselhaus SO fields have the same symmetry, it should be easy to attain both the weak and strong SO regimes, via a sensitive electrical control of the Rashba couplings (both magnitude and sign). More interesting, it may be possible to tune $\alpha=-\beta$ in some parameter range, which would lead to a complete cancellation of the Rashba and Dresselhaus terms, thus rendering the spin a conserved quantity in wurtzite wells. This could possibly extend the SO-induced spin-flip times as compared to zincblende structures, provided that cubic contributions can be neglected. Additional work is needed to investigate this exciting possibility.
	
We hope that our general effective SO Hamiltonian stimulate further theoretical studies involving quantum wires and dots in wurtzite systems; these may be relevant, for instance, to wurtzite dot qubits.        

\begin{acknowledgments}
 This work was supported by FAPESP grant No. 2016/08468-0, CNPq, PRP/USP (Q-NANO),  the National Natural Science
Foundation of China (Grant Nos.~11004120 and 11874236), and the Research Fund of Qufu Normal University. The authors acknowledge P. E. Faria Junior for useful discussions.

\end{acknowledgments}

\appendix

\renewcommand\thefigure{\thesection.\arabic{figure}} 

\section{Effective Models via Group Theory and \texorpdfstring{\kp}{kp}}
\label{app:grouptheory}

As discussed in the main text, there are many different approaches in the literature to describe the band structure of a wurtzite crystal. Indeed, its bulk Hamiltonian has been written in many different basis sets~\cite{voon-book}, which might lead to inconsistencies among the models. In the main text, we have shown a comprehensive and detailed derivation of effective models for bulk wurtzite, as well as heterostructures (quantum wells, wires, and dots). In this appendix, we complement our discussion with a full and systematic derivation of the $2\times 2$ electronic model and the $8\times 8$ Kane Hamiltonian using group theory methods combined with the \kp approach. This provides a deeper understanding of the $s$--$p_z$ mixing and justifies the choice of basis shown in Table~\ref{table:basiscc}.

First, we present an alternative derivation of the effective electron Hamiltonian ${\cal H}(k_x,k_y)$ for the conduction band of wurtzite quantum wells near the $\Gamma$ point. Next, we derive the $8\times 8$ Kane model for bulk wurtzite. More specifically, we start with the CC basis~\cite{Chuang}, listed on the second column of Table~\ref{table:basiscc2}, and then discuss the transformation to the primed basis CC$'$, third column of Table \ref{table:basiscc2}, which is the basis set we use throughout the main text (Sec.~\ref{subsec:kaneb}). As we will see below, the $S'$ and $Z'$ orbitals arise from the $s$--$p_z$ mixing discussed in Ref.~\cite{Voon}. It is important to track the effects of this hybridization to establish all matrix elements allowed by symmetry and identify systematically the approximations used to obtain the final Kane model for wurtzite crystals. 
\begin{table}[h]
	\caption{CC basis without $\ket{\nu}$ and with $\ket{\nu'}$ the $s$-$p_z$ hybridization. Normalization requires $|q_s|^2 + |q_z|^2 = 1$. The $q_s$ and $q_z$ coefficients here match those of Ref.~\cite{Voon}. The orbitals $S$ and $Z$ belong to the IRREP $\Gamma_1$ (transform as scalars), while $X$ and $Y$ combine to form $\Gamma_5$ [transform as a $(x,y)$ vector].}
	\begin{ruledtabular}
		\begin{tabular}{cccc}
			$\nu$ & $\ket{\nu}$ & $\ket{\nu'}$ & $C_{6v}$ IRREP \\ \hline
			1 & $\ket{iS\uparrow}$ & $\ket{iS'\uparrow} = q_s \ket{iS\uparrow} + i q_z\ket{Z\uparrow}$ & $\Gamma_1$ \\
			2 & $\ket{iS\downarrow}$ & $\ket{iS'\downarrow} = q_s \ket{iS\downarrow} + i q_z\ket{Z\downarrow}$ & $\Gamma_1$ \\
			3 & $-\frac{1}{\sqrt 2}\ket{X+iY\uparrow}$ & $-\frac{1}{\sqrt 2}\ket{X+iY\uparrow}$ & $\Gamma_5$ \\
			4 & $+\frac{1}{\sqrt 2}\ket{X-iY\downarrow}$ & $+\frac{1}{\sqrt 2}\ket{X-iY\downarrow}$ & $\Gamma_5$ \\
			5 & $+\frac{1}{\sqrt 2}\ket{X-iY\uparrow}$ & $+\frac{1}{\sqrt 2}\ket{X-iY\uparrow}$ & $\Gamma_5$ \\
			6 & $-\frac{1}{\sqrt 2}\ket{X+iY\downarrow}$ & $-\frac{1}{\sqrt 2}\ket{X+iY\downarrow}$ & $\Gamma_5$ \\
			7 & $\ket{Z\uparrow}$ & $\ket{Z'\uparrow} = q_s\ket{Z\uparrow} + i q_z\ket{iS\uparrow}$ & $\Gamma_1$ \\
			8 & $\ket{Z\downarrow}$ & $\ket{Z'\downarrow} = q_s\ket{Z\downarrow} + i q_z\ket{iS\downarrow}$ & $\Gamma_1$ \\
		\end{tabular}
	\end{ruledtabular}
	\label{table:basiscc2}
\end{table}

The last column in Table \ref{table:basiscc2} shows the irreducible representations (IRREPs) of the orbital part of the basis functions. Notice that, while zincblende crystals belong to the cubic $T_d$ group, the wurtzite lattice belongs to the $C_{6v}^4$. For $T_d$, the directions $x$, $y$, and $z$ are all equivalent. However, on $C_{6v}$ there is no symmetry constraint in the $z$ direction, which makes the $S$ and $Z$ orbitals equivalent in terms of their symmetries (both belong to $\Gamma_1$). This reduced symmetry leads  to the appearance of extra finite terms on wurtzite crystals. Nonetheless, this broken symmetry along the $z$ axis can be considered ``small'' (quasi-cubic approximation \cite{voon-book}). Therefore, all new terms are expected to be small or negligible. Here, however, we present all these new terms allowed by symmetry and show their \kp expressions.

\subsection{Theory of invariants}
\label{app:invariant}

The theory of invariants \cite{Bir,Winkler,Voon} allows us to search for the most general form of the Hamiltonian ${\cal H}(\bm{k})$ allowed by the symmetries of the crystal. Namely, we can write
\begin{equation}
{\cal H}(\bm{k}) = \sum_{i,j,l}  H_{i,j,l} \; k_{x}^{i} \; k_{y}^j \; k_{z}^l,
\label{Hinvariant}
\end{equation} 
where $H_{i,j,l}$ are arbitrary matrices to be found. For 2D systems one can simply drop $k_z$ and the sum over $l$. 

The group of the Schr\"odinger equation is composed by the set of symmetry operations $\{{\cal O}_i\}$ that keeps the crystal invariant. Consequently, these operations must commute with ${\cal H}(\bm{k})$, i.e., $[{\cal H}(\bm{k}),{\cal O}_i]=0$, yielding
\begin{align}
	{\cal H}(\bm{k}) &= 
	D_\psi({\cal O}_{i})
	\,
	{\cal H}\Big(D_k^{-1}({\cal O}_i)\bm{k}\Big)
	\,
	D_\psi^{-1}({\cal O}_i),
	\label{inv}
\end{align}
in which $D_\psi({\cal O}_{i})$ and $D_k({\cal O}_{i})$ are the matrix representations of ${\cal O}_{i}$ in the Hilbert and $k$ spaces, respectively. For each ${\cal O}_i$, this imposes constraints on ${\cal H}(\bm{k})$, which can be then written as a system of coupled equations solved to find the allowed terms in each $H_{i,j,l}$.

The matrix representations of $D_\psi({\cal O}_{i})$ and $D_k({\cal O}_{i})$ can be found on the Bilbao Crystallographic Server \cite{Bilbao1,Bilbao2,Bilbao3}. These can be applied to the QSYMM~\cite{qsymm} python package that solves the constraints imposed by Eq.~\eqref{inv} and returns the most general expressions allowed by symmetry for $H_{i,j,l}$.

\subsection{$2\times 2$ electronic effective model}

Let us now apply the theory of invariants to find the effective $2\times 2$ model for our two-dimensional electron gas ${\cal H}(k_x, k_y)$ [for comparison, see upper left block of Eq.~\eqref{eq:qw2d}, i.e., the one-subband case]. 

Some materials, e.g., GaN and AlGaN, crystallize into the wurtzite structure, which belongs to the $C_{6v}^4$~(or P6$_3$mc) non-
symmorphic space group~\cite{Bir}. At the $\Gamma$ point, the first conduction bands are spin degenerate. Their orbital components transform as scalars~(trivial representation), while the spinors transform as the $D_{1/2}$ double group IRREP, which is given by the generators of the SU(2). These are equivalent to the $D_{1/2}$ IRREP of the symmorphic group $C_{6v}$. For simplicity, we use the point group notation for the operations $\left\{{\cal O}_i\right\}=\left\{E,2C_6,2C_3,C_2,3\sigma_d,3\sigma_v\right\}$. In practice, it is sufficient to consider the generators $\{C_6, \sigma_d\}$, which correspond, respectively, to the six-fold rotations around the $z$ axis and the $x\rightarrow -x$ mirror operations. Additionally, we also consider time-reversal symmetry ${\cal T}$. In the Hilbert space, these read
\begin{align}
	D_\psi(C_6) =& \pm \exp\left(i \frac{\pi}{6}\sigma_z\right),
	\\
	D_\psi(\sigma_d) =& \pm \exp\left(i \frac{\pi}{2}\sigma_x\right),
	\\
	D_\psi({\cal T}) =& i \sigma_y {\cal K},
\end{align}
where ${\cal K}$ is the complex conjugate, and the $\pm$ signs refer to the $2\pi$ and $4\pi$ SU(2) rotations. On the other hand, for the k-space representation, with $\bm{k} = (k_x \quad k_y)^T$, we have 
\begin{align}
	D_k(C_{6}) =& \left[\begin{array}{cc}
		\cos\left(\frac{2\pi}{6}\right) & \sin\left(\frac{2\pi}{6}\right)\\
		-\sin\left(\frac{2\pi}{6}\right) & \cos\left(\frac{2\pi}{6}\right)
		\end{array}\right],
	\\
	D_k(\sigma_d) =& -\sigma_z,
	\\
	D_k({\cal T}) =& -1_{2\times 2}.
\end{align}
Substituting these representations into Eq.~\eqref{inv}, we find
\begin{align}
	H_{0,0}=& h_0  1_{2\times 2}
		\\
	H_{1,0}=& h_1  \sigma_y,
	\\
	H_{0,1}=& -h_1  \sigma_x,
	\\
	H_{1,1}=& 0_{2\times 2}, 
	\\
	H_{2,0}=& H_{0,2}=h_2  1_{2\times 2}
	\end{align}
with $h_0$, $h_1$, and $h_2$ arbitrary parameters. We can choose $h_0=0$, as it just results in an overall rigid shift. Moreover, we can identify $h_1 = \alpha$ and $h_2 = \hbar^2/2m^*$, yielding
\begin{equation}
	{\cal H}\left(k_{x},k_{y}\right) = \alpha\left(k_{x}\sigma_{y}-k_{y}\sigma_{x}\right)+\dfrac{\hbar^2}{2m^*}\left(k_{x}^{2}+k_{y}^{2}\right).
\label{Hfinalinv}
\end{equation}
The effective mass $m^*$ and the spin-orbit coupling $\alpha$ here remain as unknown parameters. These can be obtained by fitting \emph{ab initio} or experimental data. A more insightful approach, however, is to start from the $8\times 8$ Kane model -- derived in the next section -- and project its solutions onto the electron subspace as developed in the main text.

\subsection{$8\times 8$ Kane model: method of invariants}

We apply the method of invariants to obtain the most general $8\times 8$ Kane model considering the basis states shown in Table \ref{table:basiscc2}. Differently from the previous section, here we follow a single group formulation, which matches the CC basis, and maintain the spins along $z$ as good quantum numbers for the basis set. The representations for the generators $\{C_6, \sigma_d\}$ and the time-reversal operator ${\cal T}$ follow from the IRREPs $\Gamma_1$ and $\Gamma_5$ of the $C_{6v}$ point group. Applying the constraints of the theory of invariants, we obtain the following $8\times8$ model in the CC basis $\ket{\nu}$ 
\begin{multline}
H_{8\times8} = 
\begin{pmatrix}
{c_0} & 0 & 0 & 0 & 0 & i{c_5} & i{c_6} & 0 \\
0 & {c_0} & 0 & 0 & i{c_5} & 0 & 0 & i{c_6} \\
0 & 0 & {c_1} & 0 & 0 & 0 & 0 & 0 \\
0 & 0 & 0 & {c_1} & 0 & 0 & 0 & 0 \\
0 & -i{c_5} & 0 & 0 & {c_2} & 0 & 0 & {c_3} \\
-i{c_5} & 0 & 0 & 0 & 0 & {c_2} & {c_3} & 0 \\
-i{c_6} & 0 & 0 & 0 & 0 & {c_3} & {c_4} & 0 \\
0 & -i{c_6} & 0 & 0 & {c_3} & 0 & 0 & {c_4}
\end{pmatrix}
\\
+
\begin{pmatrix}
0 & i{c_7^-} & {c^+_{13}} & 0 & -{c^-_{14}} & 0 & 0 & -{c^-_{15}} \\
-i{c^+_7} & 0 & 0 & -{c^-_{13}} & 0 & {c^+_{14}} & {c^+_{15}} & 0 \\
{c^-_{13}} & 0 & 0 & 0 & 0 & i{c^-_8} & i{c^-_9} & 0 \\
0 & -{c^+_{13}} & 0 & 0 & -i{c^-_8} & 0 & 0 & i{c^-_9} \\
-{c^+_{14}} & 0 & 0 & +i{c^+_8} & 0 & -i{c^-_{10}} & -i{c^-_{11}} & 0 \\
0 & {c^-_{14}} & -i{c^+_8} & 0 & +i{c^+_{10}} & 0  & 0  & i{c^-_{11}} \\
0 & {c^-_{15}}  & -i{c^+_9} & 0 & +i{c^+_{11}} & 0 & 0 & i{c^-_{12}} \\
-{c^+_{15}} & 0 & 0 & -i{c^+_9} & 0 & -i{c^+_{11}} & -i{c^+_{12}} & 0
\end{pmatrix} k_\pm
\\
+
\begin{pmatrix}
0 & 0 & 0 & 0 & 0 & {c_{16}} & {c_{17}} & 0 \\
0 & 0 & 0 & 0 & {c_{16}} & 0 & 0 & {c_{17}} \\
0 & 0 & 0 & 0 & 0 & 0 & 0 & 0 \\
0 & 0 & 0 & 0 & 0 & 0 & 0 & 0 \\
0 & {c_{16}} & 0 & 0 & 0 & 0 & 0 & i{c_{18}} \\
{c_{16}} & 0 & 0 & 0 & 0 & 0 & i{c_{18}} & 0  \\
{c_{17}} & 0 & 0 & 0 & 0 & -i{c_{18}} & 0 & 0 \\
0 & {c_{17}} & 0 & 0 & -i{c_{18}} & 0 & 0 & 0
\end{pmatrix} k_z,
\label{eq:H88}
\end{multline}
where $c_n$, $n=0, \dots, 18$, are unknown coefficients allowed by symmetry, which later on will be defined in terms of the \kp parameters. The first matrix above corresponds to the $k$-independent terms, the second matrix gives the terms $c_n^\pm$ linear in $k_\pm = k_x \pm i k_y$, and the third matrix represents the $k_z$-linear terms. For simplicity, we do not show the $k_x^2$, $k_y^2$, and $k_z^2$ terms.

The model in Eq.~\eqref{eq:H88} is built solely based on the symmetries of the basis set. However, the IRREPs that define the $\ket{\nu}$ and $\ket{\nu'}$ basis sets are the same, as shown in Table \ref{table:basiscc2}. Hence, up to this point, the model above does not distinguish between different basis sets with equivalent symmetries. To proceed we must specify the basis and identify the matrix elements $c_n$ in terms of the \kp theory.

\subsection{$8\times 8$ Kane model: $ {\bf k}\cdot {\bf p}$~approach}
\label{subsec:kp}

Here we consider $H = H_0^{\rm CC} + W^{\rm CC}(\bm{k})$, with
\begin{align}
\label{eq:h0cc2}
H^{\rm CC}_0 &= \frac{p^2}{2m_0}+V(\bm{r}) + H_{{\rm so}z},
\\
\label{eq:wkcc}
W^{\rm CC}(\bm{k}) &= \cfrac{\hbar}{m_0}\bm{k} \cdot \bm{\pi} + H_{{\rm so}x}+H_{{\rm so}y},
\\
H_{{\rm so}j} &= \mathcal{C}\,\Bigg(\bm{\nabla}
V(\bm{r})\times\bm{p}\Bigg)_j\,\sigma_j,
\\
\bm{\pi} &= \bm{p} + \mathcal{C}\Bigg[ \bm{\sigma}\times\bm{\nabla}V(\bm{r})\Bigg],
\end{align}
where $\mathcal{C} = \hbar/4m_0^2c^2$ defines the SO coupling intensity. 

Below we first show how a commutator trick \cite{Voon} can be used to simplify some matrix elements and later we identify the matrix elements using the $\kp$ model above.

\subsubsection{Selection rules and commutator trick}
\label{sec:cardona}

As shown in Eq.~\eqref{eq:h0cc2}, we choose to keep only the $z$-component of $H_{{\rm so}}$ in $H^{\rm CC}_0$, such that the spin remains a good quantum number. In addition, this choice allows us to use a commutator trick introduced in Ref.~\cite{Voon} to eliminate a few matrix elements. Namely, the matrix elements of $\mathcal{C}\sigma_x V_z \equiv \mathcal{C}\sigma_x \partial_{z}V$, which can be written as
\begin{align}
	\mathcal{C}\bra{\nu_1 \uparrow} \sigma_x V_z \ket{\nu_2 \downarrow} =& \, \mathcal{C}\bra{\nu_1 \uparrow} V_z \ket{\nu_2 \uparrow},
	\label{eq:cardona1}
	\\
	\nonumber
	\mathcal{C}\bra{\nu_1 \uparrow} \sigma_x V_z \ket{\nu_2 \downarrow} =&
	-\dfrac{i}{\hbar} \mathcal{C} \bra{\nu_1\uparrow}[V, p_z]\ket{\nu_2\uparrow}
	\\
	\nonumber
	=& -\dfrac{i}{\hbar} \mathcal{C} \bra{\nu_1\uparrow}[H_0^{\rm CC}-H_{{\rm so}z}, p_z]\ket{\nu_2\uparrow}
	\\
	\nonumber
	=& -\dfrac{i}{\hbar} \mathcal{C} \Bigg[ (\varepsilon_{\nu_1\uparrow}-\varepsilon_{\nu_2\uparrow})\bra{\nu_1\uparrow}p_z\ket{\nu_2\uparrow} 
	\\
	\nonumber
	&\quad\quad\quad - \bra{\nu_1\uparrow}[H_{{\rm so}z}, p_z]\ket{\nu_2\uparrow}\Bigg]
	\\
	=& \dfrac{i}{\hbar} \mathcal{C} \bra{\nu_1\uparrow}[H_{{\rm so}z}, p_z]\ket{\nu_2\uparrow}.
	\label{eq:cardona2}
\end{align}
Here $\ket{\nu \sigma}$ are the eigenstates of $H_0^{\rm CC}$ with eigenenergies $\varepsilon_{\nu \sigma}$. Similar expressions hold for the matrix elements of $\mathcal{C}\sigma_y V_z$. In Eq.~\eqref{eq:cardona1}, we have simply acted $\sigma_x$ on the ket, while in Eq.~\eqref{eq:cardona2}, we have used the identity $\partial_z V \equiv -i[V,p_z]/\hbar \equiv -i[H_0^{\rm CC}-H_{{\rm so}z},p_z]$. The last line in~\eqref{eq:cardona2} is only valid if the eigenstates are degenerate, i.e., $\varepsilon_{\nu_1\uparrow} = \varepsilon_{\nu_2\uparrow}$. Below we show that this type of matrix element, either in the form \eqref{eq:cardona1} or \eqref{eq:cardona2}, is identically zero or negligible.

Let us first consider the matrix element in the form \eqref{eq:cardona1}. The operator $V_z$ transforms as $\Gamma_1$, while the eigenstates transform either as $\Gamma_1$ or $\Gamma_5$ (see Table \ref{table:basiscc2}). Therefore, the selection rules already dictate that the matrix element is zero if $\ket{\nu_1 \sigma_1}$ and $\ket{\nu_2 \sigma_2}$ belong to different IRREPs. 

On the other hand, if they belong to the same IRREP $\Gamma_{\nu_1} = \Gamma_{\nu_2} = \Gamma_j$ ($j=1$ or $5$), we have to argue differently, since $\Gamma_1 \otimes \Gamma_1 = \Gamma_1$, and $\Gamma_5 \otimes \Gamma_5 \supset \Gamma_1$. Consider then the form \eqref{eq:cardona2} for degenerate states ($\varepsilon_{\nu_1\uparrow} = \varepsilon_{\nu_2\uparrow}$). Notice that the operator $[H_{{\rm so}z}, p_z]$ transforms as $\Gamma_2$. The matrix element transforms as $\Gamma_j \otimes \Gamma_2 \otimes \Gamma_j$ and two cases are possible:
\begin{itemize}
	\item[(a)] $\Gamma_{\nu_1}=\Gamma_{\nu_2}=\Gamma_1$: \thinspace $\Gamma_1 \otimes \Gamma_2 \otimes \Gamma_1 = \Gamma_2$, which yields $\bra{\nu_1 \uparrow} [H_{{\rm so}z}, p_z] \ket{\nu_2 \downarrow}=0$;
	\item[(b)] $\Gamma_{\nu_1}=\Gamma_{\nu_2}=\Gamma_5$: \thinspace $\Gamma_5 \otimes \Gamma_2 \otimes \Gamma_5 \supset \Gamma_1$, which allows for $\bra{\nu_1 \uparrow} [H_{{\rm so}z}, p_z] \ket{\nu_2 \downarrow}\neq0$.
\end{itemize}
It is easy to see, though, that the nonzero matrix element in (b) results in a negligible higher order correction, since $\frac{i}{\hbar}\mathcal{C}\bra{\nu_1 \uparrow} [H_{{\rm so}z}, p_z] \ket{\nu_2 \downarrow}\propto\mathcal{C}^2$. We shall point out that this leads to the term $\alpha_2$ obtained in Ref~\cite{Paulo}.

In summary, the matrix elements of $\bra{\nu_1 \uparrow} \sigma_\mu V_z \ket{\nu_2 \downarrow}$, with $\mu=\{x,y\}$, are 
\begin{itemize}
	\item [(1)] identically zero 
	\begin{itemize}
		\item[(i)] if $\ket{\nu_1 \uparrow}$ and $\ket{\nu_2\downarrow}$ belong to different IRREPs;
		\item[(ii)] if $\varepsilon_{\nu_1\uparrow} = \varepsilon_{\nu_2\uparrow}$ and $\ket{\nu_1 \uparrow}$ and $\ket{\nu_2\downarrow}$ belong to the $\Gamma_1$ IRREP. 
	\end{itemize}
	\item [(2)] negligible ($\propto \mathcal{C}^2$)
	\begin{itemize}
		\item[(i)] if $\varepsilon_{\nu_1\uparrow} = \varepsilon_{\nu_2\uparrow}$ and $\ket{\nu_1 \uparrow}$ and $\ket{\nu_2\downarrow}$ belong to the $\Gamma_5$ IRREP.  
	\end{itemize}
\end{itemize}
Notice that, due to the spin-flip induced by the $\sigma_\mu$ operator acting on the matrix element, the states $\ket{\nu_1\uparrow}$ and $\ket{\nu_2\uparrow}$ must have the same energies, instead of $\ket{\nu_1\uparrow}$ and $\ket{\nu_2\downarrow}$.  

The selection rules above rely on the form of our Hamiltonian given in Eqs.~\eqref{eq:h0cc2} and~\eqref{eq:wkcc}. This is a partially relativistic model, accounting only for the SO correction. Additionally, one could consider the scalar relativistic corrections, namely the mass-velocity $H_{mv}=\frac{p^4}{
8 m_0^3c^2}$ and the Darwin $H_D=\frac{\hbar^2
}{8m^2c^2}\nabla^2V(\mathbf{r})$ terms. Both transform as $\Gamma_1$ and might contribute to the $k=0$ diagonal matrix elements (band edges) and to the $s$--$p_z$ hybridization [$c_6$ term in Eq.~\eqref{eq:H88}]. More importantly, these would also appear added to $H_{{\rm so}z}$ in Eq.~\eqref{eq:cardona2}, possibly affecting the selection rule (1-ii). The mass-velocity term vanishes as $[H_{mv}, p_z] = 0$. However, the Darwin term breaks the selection rule $[H_D, p_z] \neq 0$. Nonetheless, since the latter arises from the fine structure, it scales with $H_D \propto \mathcal{C}$, yielding again a negligible contribution $\mathcal{C}\bra{\nu_1 \uparrow} \sigma_x V_z \ket{\nu_2 \downarrow} = \frac{i}{\hbar}\mathcal{C}\bra{\nu_1\uparrow}[H_D, p_z]\ket{\nu_2\uparrow} \propto \mathcal{C}^2$. This correction allows for the $\gamma_1$ and $\alpha_3$ terms in Ref.~\cite{Paulo}.

\subsubsection{Matrix elements}

Next, let us follow the \kp approach and identify the leading order contributions for each $\bm{k}=0$ term in $H_{8\times8}$ from Eq.~\eqref{eq:H88}. After simplifications, these are
\begin{align}\nonumber
{c_0 } &= \expval{\dfrac{p^2}{2m_0} + V(\bm{r})}{iS\uparrow},
\\\nonumber
{c_4 } &= \expval{\dfrac{p^2}{2m_0} + V(\bm{r})}{Z\uparrow},
\\\nonumber
{ic_6} &= \mel{iS\uparrow}{\dfrac{p^2}{2m_0} + V(\bm{r})}{Z\uparrow},
\\\nonumber
{c_3 } &= +\sqrt{2} \mathcal{C}\mel{X-iY\uparrow}{V_yp_z-V_zp_y}{Z\uparrow},
\\\nonumber
{-ic_5} &= +\sqrt{2} \mathcal{C}\mel{X-iY\uparrow}{V_yp_z-V_zp_y}{iS\uparrow}.
\end{align}
Both ${c_6}$ and ${c_5}$ are nonzero due to the broken wurtzite symmetry along $z$ (in zincblende $c_6 = c_5 = 0$). It remains to define $c_1 = c_X + \delta c_X$ and $c_2 = c_X - \delta c_X$, in which
\begin{align}\nonumber
c_X &= \dfrac{1}{2} \mel{X+iY\uparrow}{\dfrac{p^2}{2m_0}+V(\bm{r})}{X+iY\uparrow},
\\\nonumber
\delta c_X &= \dfrac{1}{2}\mathcal{C}\mel{X+iY\uparrow}{V_xp_y-V_yp_x}{X+iY\uparrow}.
\end{align}
To properly use the commutator trick introduced in Eq.~\eqref{eq:cardona2}, it is important to keep track of the matrix form of $H_0^{\rm CC}$, which includes only the $z$-component of the SO coupling. Within this non-primed basis, we get
\begin{align}\nonumber
H_0^{\rm CC} = 
\begin{pmatrix}
	{c_0} & 0 & 0 & 0 & 0 & 0 & i{c_6} & 0 \\
	0 & {c_0} & 0 & 0 & 0 & 0 & 0 & i{c_6} \\
	0 & 0 & {c_1} & 0 & 0 & 0 & 0 & 0 \\
	0 & 0 & 0 & {c_1} & 0 & 0 & 0 & 0 \\
	0 & 0 & 0 & 0 & {c_2} & 0 & 0 & 0 \\
	0 & 0 & 0 & 0 & 0 & {c_2} & 0 & 0 \\
	-i{c_6} & 0 & 0 & 0 & 0 & 0 & {c_4} & 0 \\
	0 & -i{c_6} & 0 & 0 & 0 & 0 & 0 & {c_4}
\end{pmatrix}.
\end{align}
The term ${c_6}$ is the one responsible for the $s$--$p_z$ mixing \cite{Voon}, as it couples the states $\ket{S}$ and $\ket{Z}$.

The finite terms on the $k_z$ block of $H_{8\times8}$ are
\begin{align}\nonumber
	{c_{16}} &= \dfrac{\hbar}{m_0} \dfrac{\mathcal{C}}{\sqrt{2}}\mel{X-iY\uparrow}{V_y+iV_x}{iS\uparrow},
\\\nonumber
	{ic_{18}} &= \dfrac{\hbar}{m_0} \dfrac{\mathcal{C}}{\sqrt{2}}\mel{X-iY\uparrow}{V_y+iV_x}{Z\uparrow},
\\\nonumber
	c_{17} &= \dfrac{\hbar}{m_0}\mel{iS\uparrow}{p_z}{Z\uparrow} = P_1.
\end{align}
The term ${c_{16}}$ is allowed in both zincblende and wurtzite, while ${c_{18}}$ only in wurtzite. However, one typically neglects these $k$-dependent SO terms. The term ${c_{17}}$ gives us the Kane parameter $P_1$, also present in the CC original paper \cite{Chuang}.

The $k_\pm$-dependent terms of $H_{8\times 8}$ are given by
\begin{align}
ic_8 &= \dfrac{\hbar}{m_0}\dfrac{\mathcal{C}}{2} \bra{X+iY\uparrow} \sigma_y V_z \ket{X+iY\downarrow},
\\\nonumber
{ic_7} &= -i\dfrac{\hbar\mathcal{C}}{m_0}\expval{V_z}{iS\uparrow},
\\\nonumber
ic_{10} &= -i\dfrac{\hbar\mathcal{C}}{2m_0} \mel{X-iY\uparrow}{V_z}{X+iY\uparrow},
\\\nonumber
{ic_{12}} &= -i\dfrac{\hbar\mathcal{C}}{m_0}\expval{V_z}{Z\uparrow},
\\\nonumber
{c_{15}} &= -i \dfrac{\hbar\mathcal{C}}{m_0}\mel{Z\uparrow}{V_z}{iS\uparrow},
\\	\nonumber
	{ic_9} &= \dfrac{\hbar}{\sqrt{2}m_0}\Bigg[ \mel{X-iY\downarrow}{p_x}{Z\downarrow} + \mathcal{C}\mel{X-iY\downarrow}{V_y}{Z\downarrow}\Bigg],
\\
	\nonumber
	{ic_{11}} &= \dfrac{-\hbar}{\sqrt{2}m_0}\Bigg[ \mel{X+iY\downarrow}{p_x}{Z\downarrow} + \mathcal{C}\mel{X+Y\downarrow}{V_y}{Z\downarrow}\Bigg],
\\
	\nonumber
	{c_{13}} &= \dfrac{-\hbar}{\sqrt{2}m_0}\Bigg[ \mel{X-iY\downarrow}{p_x}{iS\downarrow} + \mathcal{C}\mel{X-iY\downarrow}{V_y}{iS\downarrow}\Bigg],
\\
	\nonumber
	{c_{14}} &= \dfrac{-\hbar}{\sqrt{2}m_0}\Bigg[ \mel{X+iY\downarrow}{p_x}{iS\downarrow} + \mathcal{C}\mel{X+iY\downarrow}{V_y}{iS\downarrow}\Bigg].
\end{align}
The $k$-dependent SO components in ${c_{13}}$ and ${c_{14}}$ are usually neglected, which leads to ${c_{13} = c_{14}}$ for both zinc-blend and wurtzite. Under the same approximation, ${c_7}$, $c_{10}$, ${c_{12}}$, and ${c_{15}}$ can be neglected. Terms ${c_9 \approx -c_{11}}$ can be neglected under the quasi-cubic assumption. As mentioned above, we cannot use the commutator trick to eliminate $c_7$ and $c_{12}$, since $\ket{iS\uparrow}$ and $\ket{Z\uparrow}$ are not eigenstates of $H_0^{\rm CC}$ due to the possible $s$--$p_z$ mixing introduced by $c_6$. This is a strong motivation to change to the primed basis $\ket{\nu'}$ from Table \ref{table:basiscc2}, which will not only eliminate $c_6$ through a rotation, but will also allow us to use the selection rules and the commutator trick to simplify the final model.

\subsubsection{Changing basis: $s$--$p_z$ mixing}

To eliminate the term $c_6$, the $\ket{\nu'}$ basis is defined by the coefficients $q_s$ and $q_z$ given by the eigenstates of the submatrix $\left(\begin{smallmatrix}c_0 & i c_6 \\-i c_6 & c_4\end{smallmatrix}\right)$, which read
\begin{align}
q_s \approx 1 - \dfrac{1}{2}\left(\dfrac{{c_6}}{{c_0-c_4}}\right)^2,
\quad
q_z \approx -\left(\dfrac{{c_6}}{{c_0-c_4}}\right).
\end{align}
As expected, $q_s$ is defined by the ratio between the coupling ${c_6}$ and the gap ${c_0-c_4} = {\varepsilon_{1\uparrow}-\varepsilon_{8\uparrow}}$, hence $|q_z| \ll |q_s|$. The resulting primed basis $\ket{\nu'}$ diagonalizes $H_0^{\rm CC}$,
\begin{align}\nonumber
H_0^{\prime \; \rm CC} = 
\begin{pmatrix}
{c'_0} & 0 & 0 & 0 & 0 & 0 & 0 & 0 \\
0 & {c'_0} & 0 & 0 & 0 & 0 & 0 & 0 \\
0 & 0 & {c_1} & 0 & 0 & 0 & 0 & 0 \\
0 & 0 & 0 & {c_1} & 0 & 0 & 0 & 0 \\
0 & 0 & 0 & 0 & {c_2} & 0 & 0 & 0 \\
0 & 0 & 0 & 0 & 0 & {c_2} & 0 & 0 \\
0 & 0 & 0 & 0 & 0 & 0 & {c'_4} & 0 \\
0 & 0 & 0 & 0 & 0 & 0 & 0 & {c'_4}
\end{pmatrix}.
\end{align}

Let us now show the general Hamiltonian $H'_{8\times8}$ within the $\ket{\nu'}$ basis, with all symmetry allowed terms. In the next section we will select only the relevant matrix elements to build our final model. As previously done, we break the new Hamiltonian as $H'_{8\times8} = H'_0 + H'_z k_z + H'_{\pm}k_\pm$. 

The Hamiltonian $H'_0$ reads
\begin{equation}\nonumber
H'_0 = H_0^{\prime \; \rm CC} + 
\begin{pmatrix}
0 & 0 & 0 & 0 & 0 & i {c'_5} & 0 & 0 \\
0 & 0 & 0 & 0 & i {c'_5} & 0 & 0 & 0 \\
0 & 0 & +\delta c_X & 0 & 0 & 0 & 0 & 0 \\
0 & 0 & 0 & +\delta c_X & 0 & 0 & 0 & 0 \\
0 & -i {c'_5} & 0 & 0 & -\delta c_X & 0 & 0 & {c'_3} \\
-i {c'_5} & 0 & 0 & 0 & 0 & -\delta c_X & {c'_3} & 0 \\
0 & 0 & 0 & 0 & 0 & {c'_3} & 0 & 0 \\
0 & 0 & 0 & 0 & {c'_3} & 0 & 0 & 0
\end{pmatrix},
\end{equation}
with
\begin{align}
	\nonumber
	{c_0'} &= q_s^2 c_0 + q_z^2 c_4 -2 q_s q_z c_6 = \expval{\dfrac{p^2}{2m_0} + V(\bm{r})}{iS'\uparrow},
\\
	\nonumber
	{c_4'} &= q_s^2 c_4 + q_z^2 c_0 +2 q_s q_z c_6 = \expval{\dfrac{p^2}{2m_0} + V(\bm{r})}{Z'\uparrow},
\\
	\nonumber
	{c'_3} &= q_s c_3 + q_z c_5 = +\sqrt{2} \mathcal{C}\mel{X-iY\uparrow}{V_yp_z-V_zp_y}{Z'\uparrow},
\\
	\nonumber
	{c'_5} &= q_s c_5 - q_z c_3 = i\sqrt{2}\mathcal{C}\mel{X-iY\uparrow}{V_yp_z-V_zp_y}{iS'\uparrow}.
\label{eq:c5qs}
\end{align}
All $c'_j$ matrix elements are similar to their non-primed $c_j$ counterparts, except for the the replacements $S\rightarrow S'$ and $Z\rightarrow Z'$. The same is valid for the finite $k_z$ and $k_\pm$ terms below. Indeed, for the $k_z$-linear terms, $H'_z$ keeps the same form as its non-primed counterpart in $H_{8\times8}$, with the coefficients replaced by its primed versions as
\begin{align}
	\nonumber
	{c'_{17} = c_{17}} &= \dfrac{\hbar}{m_0}\mel{iS'\uparrow}{p_z}{Z'\uparrow},
\\
	\nonumber
	{c'_{16}} = q_s{c_{16}} - q_z{c_{18}} &= \dfrac{\hbar}{m_0} \dfrac{\mathcal{C}}{\sqrt{2}}\mel{X-iY\uparrow}{V_y+iV_x}{iS'\uparrow},
\\
	\nonumber
	{c'_{18}} = q_s {c_{18}} + q_z{c_{16}} &= -i\dfrac{\hbar}{m_0} \dfrac{\mathcal{C}}{\sqrt{2}}\mel{X-iY\uparrow}{V_y+iV_x}{Z'\uparrow}.
\end{align}
As mentioned above, $c'_{16}$ and $c'_{18}$ are $k$-dependent SO terms that are usually neglected, while $c'_{17} = P_1$ is one of the usual Kane parameters.

The change of basis truly pays off due to the simplifications on $H'_{\pm}$. We can now use the commutator trick to further eliminate
\begin{align}
\nonumber
	{ic'_7} &= -i\dfrac{\hbar\mathcal{C}}{m_0}\expval{V_z}{iS'\uparrow} = 0,
\\\nonumber
	{ic'_{12}} &= -i\dfrac{\hbar\mathcal{C}}{m_0}\expval{V_z}{Z'\uparrow} = 0.
\end{align}
Therefore, we get
\begin{equation}
\nonumber
H'_\pm = 
\begin{pmatrix}
0 & 0 & {c^{\prime +}_{13}} & 0 & -{c^{\prime -}_{14}} & 0 & 0 & -{c^{\prime -}_{15}} \\
0 & 0 & 0 & -{c^{\prime -}_{13}} & 0 & {c^{\prime +}_{14}} & {c^{\prime +}_{15}} & 0 \\
{c^{\prime -}_{13}} & 0 & 0 & 0 & 0 & 0 & i{c^{\prime -}_9} & 0 \\
0 & -{c^{\prime +}_{13}} & 0 & 0 & 0 & 0 & 0 & i{c^{\prime -}_9} \\
-{c^{\prime +}_{14}} & 0 & 0 & 0 & 0 & -ic_{10}^{\prime -} & -i{c^{\prime -}_{11}} & 0 \\
0 & {c^{\prime -}_{14}} & 0 & 0 & +ic_{10}^{\prime +} & 0  & 0  & i{c^{\prime -}_{11}} \\
0 & {c^{\prime -}_{15}}  & -i{c^{\prime +}_9} & 0 & +i{c^{\prime +}_{11}} & 0 & 0 & 0 \\
-{c^{\prime +}_{15}} & 0 & 0 & -i{c^{\prime +}_9} & 0 & -i{c^{\prime +}_{11}} & 0 & 0
\end{pmatrix},
\end{equation}
with $c'_{10} = c_{10}$, and
\begin{align}
	\nonumber
	{ic'_9} &= \dfrac{\hbar}{\sqrt{2}m_0}\Bigg[ \mel{X-iY\downarrow}{p_x}{Z'\downarrow} + \mathcal{C}\mel{X-iY\downarrow}{V_y}{Z'\downarrow}\Bigg],
\\
	\nonumber
	{ic'_{11}} &= \dfrac{-\hbar}{\sqrt{2}m_0}\Bigg[ \mel{X+iY\downarrow}{p_x}{Z'\downarrow} + \mathcal{C}\mel{X+Y\downarrow}{V_y}{Z'\downarrow}\Bigg],
\\
	\nonumber
	{c'_{13}} &= \dfrac{-\hbar}{\sqrt{2}m_0}\Bigg[ \mel{X-iY\downarrow}{p_x}{iS'\downarrow} + \mathcal{C}\mel{X-iY\downarrow}{V_y}{iS'\downarrow}\Bigg],
\\
	\nonumber
	{c'_{14}} &= \dfrac{-\hbar}{\sqrt{2}m_0}\Bigg[ \mel{X+iY\downarrow}{p_x}{iS'\downarrow} + \mathcal{C}\mel{X+iY\downarrow}{V_y}{iS'\downarrow}\Bigg],
\\
	\nonumber
	{c'_{15}} &= -i \dfrac{\hbar\mathcal{C}}{m_0}\mel{Z'\uparrow}{V_z}{iS'\uparrow}.
\end{align}

\subsubsection{Approximations and final model}

Up to this point, the results above are exactly derived from the symmetry constraints and selection rules from \S\ref{sec:cardona}. In the following, we introduce the approximations that lead to the final model $H_{8\times8}^{\rm CC}$ shown in Eq.~\eqref{eq:kanebcc} of the main text.

If one is interested in the electron bands, it is usual to neglect the $k$-dependent SO terms, which only contribute in high orders to the Löwdin partitioning. This approximation eliminates $c'_{10} \approx c'_{15} \approx c'_{16} \approx c'_{18} \approx 0$. Moreover, it allow us to neglect the SO contribution in $c'_{9}$, $c'_{11}$, $c'_{13}$, $c'_{14}$, which yields $c'_{13} \approx c'_{14} \equiv -P_2/\sqrt{2}$ as a Kane parameter, while $c'_{9} \approx -c'_{11} \approx 0$ within the quasi-cubic approximation. We have already defined $c'_{17} = P_1$. For the $\bm{k}=0$ terms, it follows that $c'_{0} = E_c$, $c_X = E_v+\Delta_1$, $\delta c_X = \Delta_2$, $c'_{4} = E_v$, $c'_{3} = \sqrt{2}\Delta_3$, and $c'_{5} = -\sqrt{2}\Delta_{\rm sz}$. By setting $E_c=0$ (energy reference) and $E_v+\Delta_1+\Delta_2=-E_g$, we have $c'_{0} = 0$, $c'_1 =-E_g$, $c'_2 = -E_g - 2\Delta_2$, and $c'_{4} = -E_g -\Delta_1-\Delta_2$. Under these assumptions, the $H'_{8\times8}$ from the previous section yields our model $H^{\rm CC}_{8\times8}$ shown in Eq.~\eqref{eq:kanebcc}.

\subsubsection{Remarks on $\Delta_3$ and $\Delta_{\rm sz}$}

For most of the terms above, the difference between the primed $c'_j$ and non-primed $c_j$ matrix elements is nearly irrelevant. However, it is worth noting that this is not the case for the terms $c'_3$ and $c'_5$. 

We have shown that ${c'_3 = q_s c_3 + q_z c_5}$. Let us assume that the $s$--$p_z$ hybridization is small, it follows that $q_z \ll q_s$ and we can take $q_s \approx 1$. The dominant contribution is ${c'_3 \approx c_3}$, which gives us
\begin{align}
	\nonumber
	\Delta_3 &= \dfrac{c'_3}{\sqrt{2}} = 
	\mathcal{C}\mel{X-iY\uparrow}{V_yp_z-V_zp_y}{Z'\uparrow},
	\\
	&\approx \dfrac{c_3}{\sqrt{2}} =  \mathcal{C}\mel{X-iY\uparrow}{V_yp_z-V_zp_y}{Z\uparrow}.
\end{align}

The case for the term $c'_5 = -\sqrt{2}\Delta_{\rm sz}$ is more delicate. The $s$--$p_z$ hybridization leads to ${c'_5 = q_s c_5 - q_z c_3}$, which may be written as
\begin{multline}
	{\Delta_{\rm sz}} = -\dfrac{q_s {c_5} - q_z {c_3}}{\sqrt{2}} = -q_s\dfrac{c_5}{\sqrt{2}} + q_z{\Delta_3} =
	\\
	\nonumber
	= \mathcal{C}\Bigg[
	-iq_s \mel{X-iY\uparrow}{V_yp_z-V_zp_y}{iS\uparrow}
	\\
	+ 
	q_z \mel{X-iY\uparrow}{V_yp_z-V_zp_y}{Z\uparrow}
	\Bigg].
\end{multline}
Notice that both matrix elements above arise from the $k$-independent SO term and are nearly identical, except for the  change $\ket{iS\uparrow} \leftrightarrow \ket{Z\uparrow}$. Therefore, we may wonder which contribution prevails. One can argue that the first matrix element (arising from $c_5$, with $\ket{iS\uparrow}$) is small under the quasi-cubic approximation. However, it multiplies $q_s \approx 1$. On the other hand, the second term (arising from $\Delta_3$, with $\ket{Z\uparrow}$) is finite even in zincblende, but multiplies $|q_z| \ll |q_s|$. Thus, it is not possible to define the dominant term \textit{a priori}.

Typically, one argues in favor of the quasi-cubic approximation to parametrize $\Delta_{\rm sz} \approx q_z \Delta_3$ in terms of $\Delta_3$, using $q_z$ as a free fitting parameter. However, the analysis above shows that this is not a good and systematic approach. Instead, we consider that it is better to keep $\Delta_{\rm sz}$ itself as an independent parameter for wurtzite crystals. Notice that $\Delta_{\rm sz}$ naturally vanishes in zincblende, for which $c_5 = q_z = 0$. \\

\section{Kane model -- diagonal basis}

\label{app:diag}

Here we obtain the real band edges and the corresponding basis set $u_{io}(\bm{r})$ that diagonalizes the Hamiltonian $H_0^{\rm diag}= H_0^{\rm CC} + H_{\rm sox} + H_{\rm soy}$ at $k=0$.

\subsection{Unstrained case}

\begin{table*}[t]
	\caption{Real band edges $\lambda_i$ ($E_i$) with $i=e,A,B,C$ and corresponding eigenfunctions ${u_i \equiv u_{i\bm{0}}(\mathbf{r})}$ of $H_0$ at $\bm{k}=0$. The relevant constants $A_i$, $c_1$, $c_2$, $E_{AB}$, and $E_{AC}$  are defined in Eqs.~\eqref{eq:e77}--\eqref{eq:norm}. The eigenergy $\lambda_i$ ($E_i$) corresponds to the band edge energy with (without) the $s$--$p_z$ mixing contribution. The difference between $\lambda_i$ and $E_i$ is less than $10^{-3}$ meV.}
	\begin{ruledtabular}
		\begin{tabular}{c|c}
			$E_i$ \hfill \hfill& $u_{i}$  \\ \hline 
			$\lambda_e\simeq E_e$ \hfill \hfill & $\ket{u_1}=A_{\lambda_e}\biggl[\ket{iS^\prime\uparrow}-\left(\cfrac{c_1^2}{E_g+E_{AB}+\lambda_e}+\cfrac{c_2^2}{E_g+E_{AC}+\lambda_e}\right)i\Delta_{\rm sz}\left(\ket{X\downarrow}+i\ket{Y\downarrow}\right)$ +$i\sqrt{2}c_1c_2\Delta_{\rm sz}\left(\cfrac{1}{E_g+E_{AB}+\lambda_e}-\cfrac{1}{E_g+E_{AC}+\lambda_e}\right)\ket{Z'\uparrow}\biggl]$ \hfill \hfill   \\
			
			$\lambda_e\simeq E_e$  \hfill \hfill & $\ket{u_2}=A_{\lambda_e}\biggl[\ket{iS^\prime\downarrow}+\left(\cfrac{c_1^2}{E_g+E_{AB}+\lambda_e}+\cfrac{c_2^2}{E_g+E_{AC}+\lambda_e}\right)i\Delta_{\rm sz}\left(\ket{X\uparrow}-i\ket{Y\uparrow}\right)$ +$i\sqrt{2}c_1c_2\Delta_{\rm sz}\left(\cfrac{1}{E_g+E_{AB}+\lambda_e}-\cfrac{1}{E_g+E_{AC}+\lambda_e}\right)\ket{Z'\downarrow}\biggl]$\hfill \hfill   \\
			
			$\lambda_A\simeq E_A$  \hfill \hfill & $\ket{u_3}=-\frac{1}{\sqrt 2}\left(\ket{X\uparrow}+i\ket{Y\uparrow}\right)$ \hfill \hfill \\
			
			$\lambda_A\simeq E_A$ \hfill \hfill & $\ket{u_4}=\frac{1}{\sqrt 2}\left(\ket{X\downarrow}-i\ket{Y\downarrow}\right)$ \hfill \hfill \\
			
			$\lambda_B\simeq E_B$  \hfill \hfill & $\ket{u_5}=A_{\lambda_B}\biggl[\ket{iS^\prime\uparrow}-\left(\cfrac{c_1^2}{E_g+E_{AB}+\lambda_B}+\cfrac{c_2^2}{E_g+E_{AC}+\lambda_B}\right)i\Delta_{\rm sz}\left(\ket{X\downarrow}+i\ket{Y\downarrow}\right)$ +$i\sqrt{2}c_1c_2\Delta_{\rm sz}\left(\cfrac{1}{E_g+E_{AB}+\lambda_B}-\cfrac{1}{E_g+E_{AC}+\lambda_B}\right)\ket{Z'\uparrow}\biggl]$ \hfill \hfill   \\
			
			$\lambda_B\simeq E_B$  \hfill \hfill & $\ket{u_6}=A_{\lambda_B}\biggl[\ket{iS^\prime\downarrow}+\left(\cfrac{c_1^2}{E_g+E_{AB}+\lambda_B}+\cfrac{c_2^2}{E_g+E_{AC}+\lambda_B}\right)i\Delta_{\rm sz}\left(\ket{X\uparrow}-i\ket{Y\uparrow}\right)$ +$i\sqrt{2}c_1c_2\Delta_{\rm sz}\left(\cfrac{1}{E_g+E_{AB}+\lambda_B}-\cfrac{1}{E_g+E_{AC}+\lambda_B}\right)\ket{Z'\downarrow}\biggl]$ \hfill \hfill  \\
			
			$\lambda_C\simeq E_C$  \hfill \hfill & $\ket{u_7}=A_{\lambda_C}\biggl[\ket{iS^\prime\uparrow}-\left(\cfrac{c_1^2}{E_g+E_{AB}+\lambda_C}+\cfrac{c_2^2}{E_g+E_{AC}+\lambda_C}\right)i\Delta_{\rm sz}\left(\ket{X\downarrow}+i\ket{Y\downarrow}\right)$ +$i\sqrt{2}c_1c_2\Delta_{\rm sz}\left(\cfrac{1}{E_g+E_{AB}+\lambda_C}-\cfrac{1}{E_g+E_{AC}+\lambda_C}\right)\ket{Z'\uparrow}\biggl]$ \hfill \hfill  \\
			$\lambda_C\simeq E_C$  \hfill \hfill & $\ket{u_8}=A_{\lambda_C}\biggl[\ket{iS^\prime\downarrow}+\left(\cfrac{c_1^2}{E_g+E_{AB}+\lambda_C}+\cfrac{c_2^2}{E_g+E_{AC}+\lambda_C}\right)i\Delta_{\rm sz}\left(\ket{X\uparrow}-i\ket{Y\uparrow}\right)$ +$i\sqrt{2}c_1c_2\Delta_{\rm sz}\left(\cfrac{1}{E_g+E_{AB}+\lambda_C}-\cfrac{1}{E_g+E_{AC}+\lambda_C}\right)\ket{Z'\downarrow}\biggl]$ \hfill \hfill   \\
			
		\end{tabular}
	\end{ruledtabular}
	\label{tabel:basisdia}
\end{table*}

The actual band edges and corresponding basis functions $u_{i0}(\bm{r})$ without strain can be obtained by diagonalizing the corresponding CC-basis Kane model [Eqs.~\eqref{eq:kanebcc}] at $\bm{k}=0$. The energy differences between the band edges, shown in Fig.~\ref{fig1}(a) (neglecting $\Delta_{\rm sz}$), are given by
\begin{equation}
\label{eq:e97}
E_{AB} = \frac{1}{2}(\Delta_{\rm cr}+3\Delta_2) - \frac{1}{2}
\sqrt{(\Delta_{\rm cr}-\Delta_2)^2+8\Delta_3^2},
\end{equation}
\begin{equation}
\label{eq:e97b}
E_{AC} = \frac{1}{2}(\Delta_{\rm cr}+3\Delta_2) + \frac{1}{2}
\sqrt{(\Delta_{\rm cr}-\Delta_2)^2+8\Delta_3^2},
\end{equation}
\begin{equation}
\label{eq:e77}
E_{BC} = E_{AC} - E_{AB} =
\sqrt{(\Delta_{\rm cr}-\Delta_2)^2+8\Delta_3^2}.
\end{equation}

The diagonal basis $u_{i0}(\bm{r})$ shown in Table~\ref{tabel:basisdia} has the following normalization constants 
\begin{equation}
\label{eq:norm}
A_i=1/{\sqrt{1+\dfrac{2c_1^2\Delta_{\rm sz}^2}{(E_g+E_{AB}+E_i)^2}
		+\dfrac{2c_2^2\Delta_{\rm sz}^2}{(E_g+E_{AC}+E_i)^2}}},\;
\end{equation}
where $A_i$ ($i$=$e,A,B,$ and $C$) and
\begin{equation}
\label{eq:ab}
c_1=\frac{E_{AC}-2\Delta_2}{\sqrt{(E_{AC}-2\Delta_2)^2+
		2\Delta_3^2}},\;
c_2=\frac{\sqrt{2}\Delta_3}{\sqrt{(E_{AC}-2\Delta_2)^2+
		2\Delta_3^2}},
\end{equation}
with $c_1^2+c_2^2=1$. The eigenvalue of $H_0$ $\lambda_i$ corresponds to the energy of the real band edges.
If we focus only on the band structure, a good approximation is to neglect the $s$--$p_z$ mixing, which yields a correction to the band edges of less than 10$^{-3}$ meV compared to the real ones. By taking the conduction band as the energy origin, namely $\lambda_e=0$, we have
\begin{align}
\label{eq:lam9}
\lambda_A\simeq E_{A} = -E_g, \\ 
\label{eq:lam7}
\lambda_B\simeq E_B= -E_g-E_{AB},\\
\label{eq:lam7p}
\lambda_C\simeq E_{C}= -E_g-E_{AC},
\end{align}
which directly maps to the band description in Fig.~\ref{fig1}(a) (solid curves). Note that, to obtain the basis functions shown in Table~\ref{tabel:basisdia}, we need to perform an exact numerical calculation of the band edge energies.

The $8\times 8$ matrix Hamiltonian in the diagonal basis is given by
\begin{widetext}
	\begin{small}
		\begin{eqnarray}
		H_{8\times 8} = \cfrac{p^2}{2m_0}
		+
		\renewcommand{\arraystretch}{2.4}
		\begin{pmatrix}
		\lambda_e & i\alpha_{\rm bulk} k_- & -A_1\frac{P_2}{\sqrt{2}}k_+ & 0 & if_{\lambda_e \lambda_B}k_- & \frac{i}{2}(g_{\lambda_e \lambda_B}k_z \atop \hspace{0.4cm} + k_zg_{\lambda_e \lambda_B}) & if_{\lambda_e \lambda_C}k_- & \frac{i}{2}(g_{\lambda_e \lambda_C}k_z \atop \hspace{0.4cm} + k_zg_{\lambda_e \lambda_C}) \\
		
		-i\alpha_{\rm bulk} k_+ & \lambda_e& 0 & A_1\frac{P_2}{\sqrt{2}}k_- & \frac{i}{2}(g_{\lambda_e \lambda_B}k_z \atop \hspace{0.4cm} + k_zg_{\lambda_e \lambda_B}) & -if_{\lambda_e \lambda_B}k_+ &  \frac{i}{2}(g_{\lambda_e \lambda_C}k_z \atop \hspace{0.4cm} + k_zg_{\lambda_e \lambda_C}) & -if_{\lambda_e \lambda_C}k_+  \\
		
		-if_{\lambda_e \lambda_B}k_+ &   -\frac{i}{2}(g_{\lambda_e \lambda_B}k_z\atop \hspace{0.4cm}+ k_zg_{\lambda_e \lambda_B}) &  \lambda_A & 0 & 0 & 0 & 0 & 0 \\
		
		-\frac{i}{2}(g_{\lambda_e \lambda_B}k_z \atop \hspace{0.4cm} + k_zg_{\lambda_e \lambda_B}) & if_{\lambda_e \lambda_B}k_-  & 0 & \lambda_A & 0 & 0 & 0 & 0 \\
		
		-if_{\lambda_e \lambda_C}k_+ & -\frac{i}{2}(g_{\lambda_e \lambda_C}k_z \atop \hspace{0.4cm} + k_zg_{\lambda_e \lambda_C})  & 0 & 0 & \lambda_B & 0 & 0 & 0 \\
		
		-\frac{i}{2}(g_{\lambda_e \lambda_C}k_z \atop \hspace{0.4cm} + k_zg_{\lambda_e \lambda_C}) &  if_{\lambda_e \lambda_C}k_- & 0 & 0 & 0 & \lambda_B & 0 & 0  \\
		
		-A_1\frac{P_2}{\sqrt{2}}k_- & 0 & 0 & 0 & 0 & 0 & \lambda_C& 0     \\
		
		0 & A_1\frac{P_2}{\sqrt{2}}k_+ & 0 & 0 & 0 & 0 & 0 & \lambda_C
		\end{pmatrix}.
		\label{eq:kanebdia}
		\end{eqnarray}
	\end{small}
\end{widetext}
The coefficient $\alpha_{\rm bulk}$ is the bulk Rashba parameter,
\begin{equation}
\label{eq:alphab}
\alpha_{\rm bulk}=2A_1^2P_2\Delta_{\rm sz}\left(\frac{c_1^2}{E_g+E_{AB}+\lambda_e}+\frac{c_2^2}{E_g+E_{AC}+\lambda_e} \right).
\end{equation}
If we properly expand the expression above, we can recover the result in Eq.~\eqref{eq:rashbab}.

The parameters $f_{\lambda_e \lambda_{i=B,C}}$ and $g_{\lambda_e \lambda_{i=B,C}}$ in Eq.~\eqref{eq:kanebdia} are defined as
\begin{eqnarray}
f_{\lambda_e \lambda_i} &=& A_{\lambda_e } A_{\lambda_i}\Bigg[c_1^2\left(\frac{1}{E_g+E_{AB}+\lambda_e} + \frac{1}{E_g+E_{AB}+\lambda_i} \right) \nonumber \\
&+& c_2^2\left(\frac{1}{E_g+E_{AC}+\lambda_e} + \frac{1}{E_g+E_{AC}+\lambda_i} \right) \Bigg],
\end{eqnarray}
and 
\begin{eqnarray}
g_{\lambda_e \lambda_i} &=& \sqrt{2} A_{\lambda_e } A_{\lambda_i} c_1 c_2 P_1 \Delta_{\rm sz} \Bigg[\left(\frac{1}{E_g+E_{AB}+\lambda_i} - \frac{1}{E_g+E_{AB}+\lambda_e} \right) \nonumber \\
&-& \left(\frac{1}{E_g+E_{AC}+\lambda_i} - \frac{1}{E_g+E_{AC}+\lambda_e} \right) \Bigg].
\end{eqnarray}

\subsection{Strained case}
\label{sec:dias}
In the presence of strain, $u_{i0}$ is obtained by diagonalizing the strained CC-basis Kane Hamiltonian (see Sec.~\ref{sec:ccs}). Therefore, the diagonal basis set is strain dependent.

To construct the Kane model using the diagonal basis in the presence of strain, we need to know how the band edges change compared to the unstrained case. One can obtain these by numerically diagonalizing the Hamiltonian. Here, however, we show the analytical expressions within the approximation of neglecting the $s$--$p_z$ mixing, since its energy correction is negligibly small.
The band edge shift due to strain at the $\Gamma$ point is schenatically shown in Fig.~\ref{fig1}(a) (dashed curves), where we have,
$E^s_e= E_e+a_{c_1}\varepsilon_{zz}$+ $a_{c_2}(\varepsilon_{xx}+\varepsilon_{yy})$, $E^s_A=E_A+S_1+S_2$,  $E^s_B  =E^s_{A}-E^s_{AB}$, and $E^s_C  =E^s_{A}-E^s_{AC}$.
The energy differences between the valence bands in the presence of strain $E^s_{AB}$, $E^s_{AC}$, and $E^s_{BC}$ read
\begin{align}
\label{eq:e97w}
E^s_{AB} &= \frac{1}{2}(\Delta_{\rm cr}+3\Delta_2+S_2) - \frac{1}{2}
\sqrt{(\Delta_{\rm cr}-\Delta_2+S_2)^2+8\Delta_3^2}, \\
\label{eq:e97bw}
E^s_{AC} &= \frac{1}{2}(\Delta_{\rm cr}+3\Delta_2+S_2) + \frac{1}{2}
\sqrt{(\Delta_{\rm cr}-\Delta_2+S_2)^2+8\Delta_3^2}, \\
\label{eq:e77w}
E^s_{BC} &= E^s_{AC} - E^s_{BC} =
\sqrt{(\Delta_{\rm cr}-\Delta_2+S_2)^2+8\Delta_3^2}.
\end{align}
We can directly compare the expressions above with the unstrained case [Eqs.~\eqref{eq:e97}--\eqref{eq:e77}] and see the corrections due to strain.

The Kane Hamiltonian in the presence of strain is again similar to that of the unstrained case, with the band parameters in Eq.~\eqref{eq:kanebdia} being replaced by the strained parameters given above.
\begin{table}[H]
	\caption{Relation between the bulk band edges $E^{CC}_i$ and $ E_i$ [Fig.~\ref{fig1}(a)], and quantum well offsets $\delta^{CC}_i$ and $ \delta_i$ [Fig.~\ref{fig1}(b)], appearing in the diagonal part of the Kane Hamiltonian. The notation refers to the unstrained case; the strained case can be straightforwardly obtained. The superscripts ``$w$'' and ``$b$'' stand for well and barrier, respectively. For the expression of energy separations
		$E_{AB}$ ($ E_{AB}$) and $E_{AC}$ ($ E_{AC}$), see Eqs.~\ref{eq:eabccc}, \ref{eq:e97}--\ref{eq:e77}, and \ref{eq:lam9}--\ref{eq:lam7p}.}
	
	\begin{ruledtabular}
		\begin{tabular}{lllc}
			$ E_e^{CC}=E_e$ & $ E_A^{CC}=E_A$ & \hspace{-1.85cm} $ E_g^{CC}=E_g$ \\
			$ E_B^{CC}=E_B+E_{AB}- E_{AB}$ & $ E_C^{CC}= E_C+E_{AC} - E_{AC}$ \\ \hline \\
			
			$ \delta_e^{CC}=\delta_e$ &  $ \delta_B^{CC}=\delta_B+(E^w_{AB}- E^w_{AB})-(E^b_{AB}- E^b_{AB})$ \\
			$ \delta_A^{CC}=\delta_A$  & $ \delta_C^{CC}=\delta_C+(E^w_{AC}- E^w_{AC})-(E^b_{AC}- E^b_{AC})$  \\
		\end{tabular}
	\end{ruledtabular}
	\label{table:bandrelation}
\end{table}


\section{Folding down approach}
\label{app:folding}
For an arbitrary $n \times n$ matrix Hamiltonian, one can always write the corresponding Schr\"odinger equation in the compact form
\begin{equation}
\label{eq:schr}
\begin{pmatrix}
H_{P} & H_{PQ} \\ H_{PQ}^{\dagger} & H_{Q}
\end{pmatrix}
\begin{pmatrix}
\Phi_P \\ \Phi_P
\end{pmatrix}
= E
\begin{pmatrix}
\Phi_P \\ \Phi_Q
\end{pmatrix},
\end{equation}
where $H_P$ and $H_Q$ are two subsets of the original matrix, $H_{PQ}$ describes the coupling between these subspaces, and  $\Phi_P$ and $\Phi_Q$ are the corresponding eigenspinors.
Here, as shown below, we are interested in the subspace $P$.

We can use the so-called folding down approach to rewrite Eq.~\eqref{eq:schr} as
\begin{equation}
\label{eq:phic}
\mathcal{H}_{P}(E)\Phi_P=E\Phi_P,  
\end{equation}
with 
\begin{equation}
\mathcal{H}_{P}(E)=\left[H_{P}+ H_{PQ}(E-H_Q)^{-1}H_{PQ}^{\dagger}\right].
\end{equation}
Note that $\mathcal{H}_{P}(E)$ is energy-dependent, which makes Eq.~\eqref{eq:phic} not a \emph{real} Schr\"odinger-type equation.

To ensure norm conservation, we build a new spinor $\Phi_P^\prime$,
\begin{equation}
\label{eq:wfrenormp}
\Phi_P^\prime = \Omega \Phi_P, \quad
\Omega = \sqrt{1+H_{PQ}
	\frac{1}{(E-H_Q)^2}H_{PQ}^\dagger}.
\end{equation}
By inserting $\Phi_P^{\prime}$ into Eq.~\eqref{eq:phic} and
multiplying the resulting equation from the left by $\Omega^{-1}$, we
arrive at
\begin{equation}
\label{eq:hcrenormp}
{\Omega}^{-1}\mathcal{H}_{P}(E){\Omega}^{-1}\Phi_P^\prime=E{\Omega}^{-2}
\Phi_P^\prime.
\end{equation}
From Eq.~\eqref{eq:hcrenormp} we obtain 
\begin{equation}
\label{eq:hcrenormp2}
\mathcal{ H}_{P}^\prime(E)\Phi_P^\prime=E\Phi_P^\prime,
\end{equation}
with
\begin{equation}
\mathcal{H}_{P}^\prime(E)=[{\Omega}^{-1}\mathcal{H}_{P}(E){\Omega}^{-1}+E(I-{\Omega}^{-2})],
\end{equation}
where $I$ is the identity matrix, which has the same dimension of the subspace $P$.
From Eq.~\eqref{eq:hcrenormp2}, it is possible to arrive at a real Schr\"odinger-type equation, i.e., $\mathcal{H}_{P}^\prime(E)\rightarrow\mathcal{H}_{P}^\prime$ energy independent, by performing a power expansion of $(E-H_Q)^{-1}$ up to the second order in the energy $E$. 

By using the procedure described above, we obtain exactly the same result as in Eqs.~\eqref{eq:well3}--\eqref{eq:hsowell}.


\section{\texorpdfstring{\kp}{kp} interaction within the valence band subspace}

\label{app:hvq}
The 6$\times$6 Hamiltonian (CC basis) for the $\bm{p}$ valence band including the \kp interaction within its subspace is given by
\begin{widetext}
\begin{small}
\begin{equation}
  \label{eq:hvq}
  H_v=\frac{p^2}{2m_0}+
 \renewcommand{\arraystretch}{2.2}
 \begin{pmatrix}
      - E_g+ V_A(z) & 0 & 0 & 0 & \frac{iQ}{\sqrt{2}}k_- & 0 \\  
    0  & - E_g+ V_A(z) & 0 & 0 & 0 & -\frac{iQ}{\sqrt{2}}k_+   \\ 
    0  & 0 & - E_g- E_{AB}+V_B(z) & 0 & -\frac{iQ}{\sqrt{2}}k_+   & \sqrt{2}\Delta_3  \\ 
    0 & 0 & 0 & - E_g- E_{AB}+V_B(z) &  \sqrt{2}\Delta_3 & \frac{iQ}{\sqrt{2}}k_+   \\
   -\frac{iQ}{\sqrt{2}}k_+  &  0 &\frac{iQ}{\sqrt{2}}k_-& \sqrt{2}\Delta_3  & - E_g- E_{AC}+V_{C}(z) & 0\\
  0 & \frac{iQ}{\sqrt{2}}k_- &\sqrt{2}\Delta_3& -\frac{iQ}{\sqrt{2}}k_+ &0 &  - E_g- E_{AC}+V_{C}(z) \\
  \end{pmatrix},
\end{equation}
\end{small}
\end{widetext}
with $Q$ being defined as  $Q=-(i\hbar/m_0)\bra{Z^\prime}p_x\ket{X}=-(i\hbar/m_0)\bra{Z^\prime}p_y\ket{Y}$.
The additional Rashba terms arising  from  the \kp interaction inside the valence band subspace read
\begin{eqnarray}
 H_R^Q= \eta_Q(z)(\sigma_x k_y - \sigma_y k_x),
\end{eqnarray}
with
\begin{eqnarray}
\nonumber  \eta_{Q}(z)&=&\cfrac{P_1Q\Delta_{\rm sz}}{( E_g+2\Delta_2)^2( E_g+\Delta_1+\Delta_2)}\cfrac{d  V_B(z)}{dz}\\ \nonumber
  &+&\cfrac{P_1Q\Delta_{\rm sz}}{( E_g+2\Delta_2)( E_g+\Delta_1+\Delta_2)^2}\cfrac{d  V_C(z)}{dz}\\
&+& \cfrac{4\Delta^2_{sz}\Delta_3Q}{( E_g+2\Delta_2)^2( E_g+\Delta_1+\Delta_2)},
\end{eqnarray}
where one can see that these terms also depend on the $s$--$p_z$ mixing, i.e., $\Delta_{\rm sz}$. The first two terms in the equation above describe the usual Rashba coupling. By assuming that the Kane parameters $P_2$ and $Q$ are comparable, we find that the strength of these terms is around one tenth of that of $\eta_H$ and $ \eta_w$ discussed in the main text [see Eqs.~\eqref{eq:etahcc} and \eqref{eq:etawcc}].
The last term contributes to the bulk Rashba term and has a negligible value when compared to~\eqref{eq:rashbab}.

In addition, we shall point out that here we also obtain
the Dresselhaus term,
\begin{eqnarray}
\label{app:dreq}
H_D=\beta_D(bk_z^2-k_{\|}^2)(\sigma_x k_y - \sigma_y k_x),
\end{eqnarray}
with the coefficients $\beta_D\equiv \beta_{D1}+\beta_{D2}$ and $b$ defined as
\begin{eqnarray}
\label{app:bd1}
\nonumber \beta_{D1}=-\frac{P^3_2\Delta_{\rm sz}}{ E_g + E_{AB}}\left(\frac{1}{ E^2_g} +\frac{1}{\left( E_g+ E_{AB}\right)^2}
+ \frac{1}{\left( E_g+ E_{AC}\right)^2} \right),\\
\end{eqnarray}
\begin{eqnarray}
\label{app:bd2}
\beta_{D2}=\frac{P_2Q\left(Q\Delta_{\rm sz}+P_2\Delta_3\right)}{\left(E_g + E_{AB}\right)\left( E_g + E_{AC}\right)}\left(\frac{1}{ E_g}+\frac{1}{ E_g+ E_{AB}}\right),
\end{eqnarray}
and
\begin{eqnarray}
\label{app:bdb}
\nonumber b=\frac{2P_1^2\Delta_3}{\beta_D \left( E_g+ E_{AB}\right)\left( E_g+ E_{AC}\right)}\left(\frac{Q\Delta_{3}-P_2\Delta_{\rm sz}}{ E_g+ E_{AC}}-\frac{P_2\Delta_{\rm sz}}{ E_g+ E_{AB}}\right).\\
\end{eqnarray}
The constants $\beta_{D1}$ and $\beta_{D2}$ describe the bulk Dresselhaus coefficients. The former arises entirely from the $s$--$p_z$ mixing and does not depend on $Q$ and the latter is determined by $Q$. It is worth mentioning that the $s$--$p_z$ mixing
also partially contributes to $\beta_{D2}$ [see Eq.~\eqref{app:bd2}].
The parameter $b$ implies the nonequivalence between the $c$ axis orientated $z$ direction and the $x$-$y$ plane.

We must emphasize that the Dresselhaus term here was obtained within the eight-band model ($\bm{s}$-conduction
and $\bm{p}$-valence bands).
This is in contrast to the zincblende structure, in which the Dresselhaus term is associated with the
coupling between the $\bm{p}$-valence and $\bm{p}$-conduction bands.
On the other hand, from Eqs.~\eqref{app:bd1}--\eqref{app:bdb} we evaluate $\beta_D\sim 0.08$ meV$\cdot$\AA$^3$ and $b\sim 0.01$, whose values are much smaller than the semi-empirical values $\beta_D\sim 0.32$ meV$\cdot$\AA$^3$ and $b\sim 4.0$~\cite{Fu}.   
Further studies to obtain the full expression of the Dresselhaus term accounting for the other remote bands are needed.
As mentioned in the main text, in this work we add the Dresselhaus coupling by hand into our self-consistent simulation and
treat the coefficients $\beta_D$ and $b$ in Eq.\ \eqref{app:dreq} as semi-empirical parameters.

\section{Total Hartree potential} 
\label{app:potential}

The total Hartree potential, given in Eq.~\eqref{eq:VHartree} of the main text, has several contributions (these arise from different charge densities): $V_{\rm elect}$, $V_{\rm int}$, $V_{\rm d}$, and $V_{\rm g}$. Here we explicitly show how to determine each one of these terms from the Poisson equation.

\subsection{Pure Hartree potential: $V_{\rm elect}$}

The pure Hartree potential $V_{\rm elect}$ is obtained by solving  
\begin{equation}
\label{eq:PoissonH}
\dfrac{d^2}{dz^2}V_{\rm elect}(z)=\dfrac{e}{\epsilon_0 \epsilon_r}\rho_e(z),
\end{equation}
together with the Dirichlet boundary conditions $V_{\rm elect}(\pm L)=0$, where $\pm L$ are the end points of our system (see Fig.~\ref{fig:pot}) and $e>0$ is the elementary charge. The parameter $\rho_e(z)$ corresponds to the electronic charge density and reads
\begin{equation}
\label{eq:density1}
\rho_e(z)=-\dfrac{2e}{A}\sum_{\nu,k_{\|}}|e^{i\mathbf{k}_{\|}\cdot \mathbf{r}}\psi_{\nu}(z)|^2f\left(\mathcal{E}_{k_{\|}\nu}\right),
\end{equation}
with $A$ the area of the electron gas in the $xy$-plane (normalizing constant) and $f\left(\mathcal{E}_{k{\|}\nu}\right)$ the Fermi-Dirac distribution. More explicitly,
\begin{equation}
\label{eq:density2}
\rho_e(z)=-\dfrac{em^*}{ \pi \hbar^2}k_B T\sum_{\nu}|\psi_{\nu}(z)|^2\ln{\left[1+e^{(\mu-\varepsilon_{\nu})/k_B T}\right],}
\end{equation}
in which $k_B$ is the Boltzmann constant, $T$ is the temperature, and $\mu$ is the chemical potential. 

We shall point that $\rho_e(z)$ in Eq.~\eqref{eq:PoissonH} depends on the wave functions $\psi_{\nu}(z)$ [see Eq.~\eqref{eq:density2}], which were obtained by solving the Schr\"odinger equation for the quantum well~\eqref{eq:hqwcc}. This equation depends on the total potential $V_e(z)$, which in turn has $V_{\rm elect}(z)$ as one of its contributions. Hence, to determine $\psi_{\nu}(z)$ and $\rho_e(z)$, we self-consistenly solve Schr\"odinger and Poisson's equations for the total charge density (see next sections for the other charge density contributions).

\subsection{Internal potential: $V_{\rm int}$}

The internal potential $V_{\rm int}$ due to the built-in electric field is written as
\begin{equation}
V_{\rm int}(z)=e\int_{-L}^z E_{\rm int}(z') dz',
\end{equation}
where $E_{\rm int}(z')$ can be either $E_w$ or $E_b$ given in  Eqs.~\eqref{eq:eboun1} (periodic boundary conditions) or ~\eqref{eq:eboun2} (neutral surface charges). The solutions in terms of the fields read

\begin{align}
V_{\rm int}(z)=\left\{\begin{array}{rll}
& eE_b\left(z+L\right), & -L\leq z\leq -{L_w}/{2} \\
& eE_b\left(L-\dfrac{L_w}{2}\right)+eE_w\left(z+\dfrac{L_w}{2}\right), & -{L_w}/{2}\leq z\leq {L_w}/{2} \\
& eE_b\left(z+L-L_w\right)+eE_wL_w. & {L_w}/{2}\leq z\leq L \\
\end{array}\right.
\end{align}
\setcounter{figure}{0} 
\begin{figure}[H]
	\includegraphics[width=8.5cm]{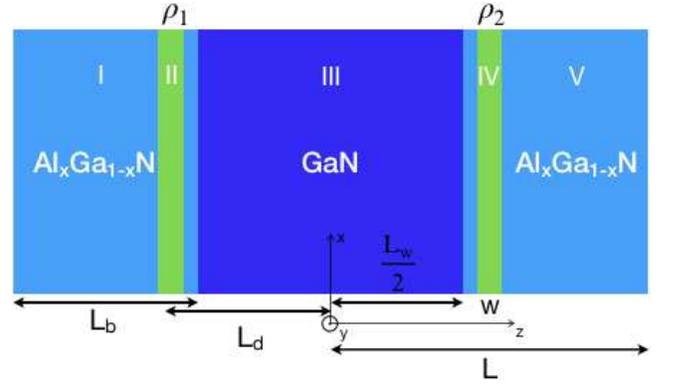}
	\caption{Schematic of a GaN/Al$_x$Ga$_{1-x}$N quantum well grown along the $z||[0001]$ direction. The well region and the two symmetric barriers have widths $L_w=10$~nm and $L_b=7$~nm, respectively. The gray regions correspond to two doping layers of densities $\rho_1$ and $\rho_2$ and width $w$, symmetrically located at $L_d=6$~nm from the center of the well. The total width of the system is $2L$.}
	\label{fig:pot}
\end{figure}

\subsection{Doping + external gate potentials: $V_{\rm d}+V_{\rm g}$}

These two contributions can be obtained by solving the Poisson equation $+$ boundary conditions in each region (I--V) of our system (Fig.~\ref{fig:pot})~\cite{Esmerindo2}. 

For the doping potential $V_d$ we have
\begin{align}
\dfrac{d^2}{dz^2}V_{\rm d}(z)=\dfrac{e^2}{\epsilon_0\epsilon_r}\left\{\begin{array}{rll}
0,& -L\leq z\leq -L_d & \hbox{(I)} \\
\rho_1,& -L_d\leq z\leq -L_d+w & \hbox{(II)} \\
0,& -L_d+w\leq z\leq L_d-w & \hbox{(III)} \\
\rho_2,& L_d-w\leq z\leq L_d & \hbox{(IV)} \\
0,& L_d\leq z\leq L & \hbox{(V)} \\
\end{array}\right.
\end{align}
where $\rho_{1,2}$ are the doping densities. The solutions to the equations above are given by
\begin{align}
V_{\rm d}(z)=\left\{\begin{array}{rll}
& a_1 z+a_2, & \hbox{(I)} \\
& \dfrac{e^2\rho_1}{2\epsilon_0\epsilon_r}z + a_3 z + a_4, & \hbox{(II)} \\
& a_5 z + a_6, & \hbox{(III)} \\
& \dfrac{e^2\rho_2}{2\epsilon_0\epsilon_r}z + a_7 z + a_8, & \hbox{(IV)} \\
& a_9 z + a_{10}. & \hbox{(V)} \\
\end{array}\right.
\end{align}
The coefficients $a_i$, $i=1,\ldots,10$, are obtained by imposing the continuity of $V_{\rm d}$ and its derivative. In addition, we consider the Dirichlet boundary conditions $V_{\rm d}(\pm L)=0$. The explicit expressions for these constants can be found in Appendix B of  Ref.~\cite{Esmerindo2}.

For the external gate potential $V_{\rm g}$, we solve 
\begin{equation}
\dfrac{d^2}{dz^2}V_{\rm g}(z)=0, \quad -L\leq z\leq L
\end{equation}
with the Dirichlet boundary conditions $V_{\rm g}(-L)=V_1$ and $V_{\rm g}(L)=V_2$, where $V_{1,2}$ are the external gates at the end points $\pm L$. We then obtain
\begin{align}
V_{\rm g}(z)= -\dfrac{(V_1-V_2)}{2L} z+ \dfrac{(V_1+V_2)}{2}.   
\end{align}




\begin{thebibliography}{100}
\bibitem{footnote-PT} This procedure is equivalent to including only first order terms in the L\"owdin expansion~\cite{Winkler}.
\bibitem{footnote-brashba} Since ${V}_{e}={V}_{H}+\delta_eh_w(z)$ and ${V}_{B}={V}_{H}- \delta_Bh_w(z)$ (see Sec.~\ref{sec:kanewell}), we have ${V}_{e}-{V}_{B}=\left(\delta_e+ \delta_B\right)h_w(z)$. Therefore, for $|z|< L_w/2$ [$h_w(z)=0$], $\eta(z)=\alpha_{\rm bulk}\rm{(well)}={2P_2\Delta_{\rm sz}}/{(E_g+2\Delta_2)}$; for  $|z|> L_w/2$ [$h_w(z)=1$], $\eta(z)=[{2P_2\Delta_{\rm sz}}/{( E_g+2\Delta_2)}] [1-(\delta_e+ \delta_B)/{(E_g+2\Delta_2)}]$, which is exactly the Taylor expansion of $\alpha_{\rm bulk}(well)={2P_2\Delta_{\rm sz}}/{( E_g+2\Delta_2+\delta_e +\delta_B)}$, up to the second order in $(\delta_e+\delta_B)/(E_g+2\Delta_2)$.
\bibitem{mixing} The $s$--$p_z$ mixing modifies the $S$ (conduction subband) and the $Z$ (valence band) wave functions, which can be written as $\ket{S}\to \ket{S'}=q_s\ket{S}+q_z\ket{Z}$ and $\ket{Z}\to \ket{Z'}=q_s\ket{Z}-q_z\ket{S}$, with $q_{s}^{2}+q_{z}^{2}=1$~\cite{Voon}. Based on the new wave functions and the  definition of $\Delta_{sz}$ by Eq.~\eqref{eq:deltasz}, one can straightforwardly obtain $\Delta_{\rm sz}=q_z\Delta_3$. Here we consider $q_z=0.1$.
\end{thebibliography}
\end{document}